\documentclass[
  aps,
  prapplied,
  reprint,
  amsmath,
  amssymb,
  superscriptaddress,
  longbibliography,
  floatfix
]{revtex4-2}
\pdfoutput=1

\usepackage{graphicx}
\usepackage{booktabs}
\makeatletter
\@ifl@t@r\fmtversion{2024/06/01}{%
  \usepackage{array}[=2016-10-06]%
}{%
  \usepackage{array}%
}
\makeatother
\usepackage{tabularx}

\usepackage{xurl}
\usepackage[hidelinks]{hyperref}
\usepackage{tikz}
\usepackage{pgfplots}
\pgfplotsset{compat=1.16}
\usetikzlibrary{arrows.meta,calc,decorations.pathreplacing}

\begin{document}

\title{Fixed-order postselected CHSH reference acquisition for
parity-constrained spatial-mode qubits on a commercial cloud photonic
processor}

\author{Emma Tully}
\affiliation{Recognition Physics Institute, Austin, TX 78701, USA}

\author{Jonathan Washburn}
\affiliation{Recognition Physics Institute, Austin, TX 78701, USA}

\author{Megan Simons}
\email{msimons@recognitionphysics.org}
\affiliation{Recognition Physics Institute, Austin, TX 78701, USA}

\begin{abstract}
We report a fixed-order Clauser--Horne--Shimony--Holt (CHSH)
acquisition and reporting protocol for correlations between two encoded
photonic qubits on Quandela's commercial, cloud-accessible Belenos
processor, executed end-to-end by external users through the public
cloud interface.  Each logical qubit is a two-dimensional spatial-mode
subspace embedded in the seven-dimensional zero-sum sector of an
eight-mode single-photon register, and a target postselected
linear-optical controlled-$Z$ (ideal success probability $1/9$)
couples the registers on $16$ of $24$ modes.  The primary quantity is the operational CHSH score $S$ on accepted
logical coincidences.  The primary analysis treats each complete
four-setting pass as the experimental unit.  Eight
sequential four-setting passes, acquired as one same-day, dependent
sequence across two sessions, each gave a raw operational score above
$2$; the session means were $2.40$ and $2.58$ with sample
standard deviations $0.15$ and $0.03$, and the eight passes showed
strong excess dispersion relative to
their conditional shot-noise errors ($Q/\nu = 6.1$, $\nu=7$), so no single
count-pooled shot-noise uncertainty is a platform precision.  We
therefore treat the raw per-pass scores and their between-pass
dispersion as the primary result and quote count-pooled quantities only as
secondary descriptors: $S_{\rm count} = 2.485 \pm 0.019$ from the raw
counts and a fixed-ratio efficiency-reweighted model scenario
$S^{\rm rw}_{\rm count} = 2.380 \pm 0.021$, whose weakest reweighted
pass ($2.040 \pm 0.062$) overlapped $2$ within $1\sigma$.  The reweighting
is an archived-metadata model scenario, not a corrected platform
score: it uses session-specific start-of-session port-transmittance
ratios whose temporal stability was not re-measured, and an
\emph{ad hoc} stress scan over equal-session ratios
$\kappa\in[1.2,1.8]$ (not a calibrated uncertainty band) spans
$2.341$--$2.441$, kept separate from the shot noise.
Setting order was fixed, the executed compiled mapping was not
returned, and residual remote-setting marginal differences remain;
consequently, the data support an operational reference acquisition
rather than an entanglement-witness or cross-platform benchmarking
claim.  The parity-check terminology labels the encoding subspace; no
syndrome measurement was performed.  Complete count records, job
identifiers, circuit-construction code, and analysis are openly
archived with content hashes for the submitted targets.
\end{abstract}

\maketitle

\section{Introduction}
\label{sec:intro}

Commercial cloud photonic processors can expose multimode
linear-optical hardware through public software interfaces while
abstracting portions of the physical implementation from the user.
On the platform studied here, the server-side compiled physical
mapping is not returned.  Characterizing such platforms therefore
requires protocols that separate intended target-circuit performance
from postselection, readout-efficiency imbalance, calibration drift
between acquisitions, and fixed acquisition order, while retaining
enough provenance that an independent user can repeat the comparison.
Cloud quantum processors have previously been used to violate
Mermin-type inequalities on bare superconducting physical
qubits~\cite{alsina2016}, while CHSH correlations with dual-rail
photonic qubits have been used in a certified-randomness experiment on
a compact photonic device~\cite{fyrillas2024}.  Quandela has also
demonstrated general-purpose cloud-accessible photonic quantum
computing on Ascella~\cite{maring2024}.  Here we study encoded
spatial-mode qubits on the 24-mode Belenos
processor~\cite{quandela_belenos,quandela_catalog}.  Accordingly, raw $S$
values are not directly comparable across dual-rail, neutral-rail,
and bare-qubit implementations: each encoding carries a different
postselection branch, mode overhead, and detection-efficiency profile,
so a raw postselected $S$ conflates hardware quality with the encoding
and postselection choices and is not a platform-independent figure of
merit across these encodings and access models.

We report an auditable user-level reference acquisition of
that kind: a postselected CHSH score between two qubits encoded in
classical-parity-check subspaces of single-photon spatial-mode
registers, obtained entirely as external users of a commercial cloud
photonic processor, with no custom hardware and no laboratory
infrastructure.  Each logical qubit occupies a two-dimensional
neutral-rail subspace of an eight-mode block; a target postselected
linear-optical controlled-$Z$ couples the registers; and the CHSH
score is assembled from archived four-way coincidence counts together
with diagnostics for readout imbalance, between-pass heterogeneity,
remote-setting marginal dependence, and acquisition-order confounding.
A companion study characterized the single-block implementation of the
zero-sum and nested parity subspaces on the same processor, realizing
the zero-sum neutral subspace as the classical $[8,7,2]$
single-parity-check code at dimension fraction $7/8$, analogous to the
classical code rate~\cite{q3paper,macwilliams1977}.
The present work uses that zero-sum rail construction but analyzes a
distinct two-photon data set.

Bell-inequality violations between logical qubits were previously
reported in bosonic~\cite{cai2024} and
superconducting~\cite{hetenyi2024} encoded systems.
More broadly, recent hardware advances have demonstrated
below-threshold quantum error correction, beyond-break-even
computation with many encoded logical qubits, and scaling and
networking of modular photonic
architectures~\cite{acharya2025,quantinuum2026,xanadu2025}.
We provide an
externally executable reference acquisition and reporting protocol
that combines encoded two-qubit photonic operations with count-level
provenance, explicit efficiency and temporal diagnostics, and an
archived analysis pipeline suitable for repetition on compatible
multimode photonic platforms.  The protocol pairs a fixed encoded
two-qubit CHSH acquisition with a heterogeneity-aware analysis
(Sec.~\ref{sec:results}) that distinguishes conditional counting
uncertainty from between-pass dispersion and shows how naive
shot-noise pooling can understate the relevant pass-level uncertainty, together with an
efficiency-reweighted analysis of the dominant identified systematic.

We observed a postselected CHSH score above $2$ in each pass of a
single same-day sequence: eight consecutive, separately complete CHSH
estimates, acquired in two sessions under different platform
calibrations, each produced a raw score $S>2$; the weakest exceeded
$2$ by $3.5$ conditional shot-noise standard deviations.  These eight
passes form one dependent, fixed-order sequence rather than
order-independent repetitions, so we do not assert repeatability across
independent acquisitions.  Because the measurement passes are
statistically heterogeneous, we treat the pass as the analysis
unit and report the between-pass spread alongside every count-pooled
descriptor, with random-effects constructions confined to sensitivity
analyses rather than a single significance figure
(Sec.~\ref{sec:results}).

The target entangling circuit
is postselected (Ralph--Langford--Bell--White
type~\cite{ralph2002,obrien2003}, ideal target success probability
$1/9$), and both qubits are prepared and read out on a single
processor.  The $[8,7,2]$ and ``parity-check'' terminology labels the
encoding subspace of Eqs.~\eqref{eq:logical}--\eqref{eq:qL}; no
zero-sum syndrome is measured in these runs.  Because the provider's compiler does not return
the physical mapping, the executed circuits are only assumed to
preserve the submitted-target factorization; the remote-setting
marginal diagnostics of Sec.~\ref{sec:results}
(Table~\ref{tab:nosig}) do not establish that assumption.
The result is therefore a fixed-order postselected CHSH-score
reference acquisition, not an entanglement-witness or loophole-free
Bell-test claim.  Ref.~\cite{q3paper} provides encoding vocabulary
and platform context; the CHSH estimate is derived entirely from the
distinct two-photon data analyzed here.

\section{Encoding and controlled-\texorpdfstring{$Z$}{Z} gate}
\label{sec:encoding}

\textit{Logical qubits.}  Throughout, ``logical'' denotes the
encoded spatial-mode degree of freedom selected by the parity
constraint below, as distinct from an individual optical mode; it
does not imply quantum error correction.  Each logical qubit
occupies one eight-mode block with modes indexed by the eight
vertices of the cube $Q_3$ (the three-dimensional hypercube graph on
the $3$-bit strings), following Ref.~\cite{q3paper}; the cube
indexing is a labeling convenience only and is not used by the gate,
CHSH analysis, or postselection below.  Within a block,
the logical basis states are dual-rail vectors whose rails are
two-mode neutral states,
\begin{equation}
\label{eq:logical}
  |0\rangle_L = \tfrac{1}{\sqrt{2}}(e_0 - e_1), \qquad
  |1\rangle_L = \tfrac{1}{\sqrt{2}}(e_2 - e_3),
\end{equation}
where $e_k$ is the standard basis vector of $\mathbb{C}^8$
representing a single photon in mode $k$ of the block; we identify
each single-photon ket with its amplitude vector, so that
$|0\rangle_L$ and $|1\rangle_L$ are the $\mathbb{C}^8$ vectors
written on the right and each logical qubit is a first-quantized
single-photon spatial-mode (path/mode-qudit)
state~\cite{erhard2020}.  Writing a general block amplitude as
$v = \sum_k v_k\,e_k \in \mathbb{C}^8$, with components
$v_k \in \mathbb{C}$, both rails satisfy the single global zero-sum
constraint $\sum_{i=0}^{7} v_i = 0$ summed over all eight block
components (not a support-local parity rule), so arbitrary logical
superpositions $\alpha|0\rangle_L + \beta|1\rangle_L$ lie in the
seven-dimensional zero-sum (``neutral'') subspace
$\mathcal{N} = \{\,v \in \mathbb{C}^8 : \sum_i v_i = 0\,\}$
defined by this single parity check.  The logical qubit itself is
the two-dimensional subspace
\begin{equation}
\label{eq:qL}
  \mathcal{Q}_L = \mathrm{span}\{|0\rangle_L, |1\rangle_L\}
  \;\subset\; \mathcal{N} \;\subset\; \mathbb{C}^8 ,
\end{equation}
so $\mathcal{N}$ is the parity-check sector that contains
the logical subspace, not the qubit space itself.  The zero-sum
hyperplane $\mathcal{N}$ is the complex-linear analogue of the
classical $[8,7,2]$ single-parity-check code, with dimension fraction
$\dim\mathcal{N}/8 = 7/8$ analogous to the classical code rate; here
$d = 2$ is the minimum support weight of the amplitude vectors, not a
quantum-code distance~\cite{macwilliams1977}.  The constraint can flag population
in the uniform-sum direction when that syndrome is measured, but it
does not correct errors or protect against photon loss.

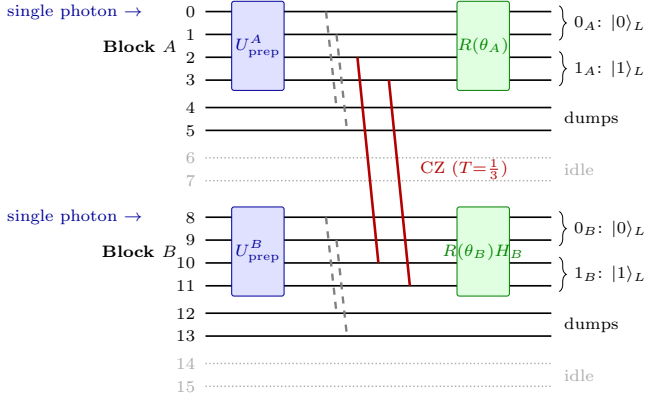
\begin{figure}[tbp]
\centering
\resizebox{\columnwidth}{!}{%
\begin{tikzpicture}[
  x=0.78cm, y=0.34cm,
  mode/.style={line width=0.7pt},
  idle/.style={line width=0.6pt, densely dotted, gray!65},
  bs/.style={line width=1.1pt, red!70!black},
  dump/.style={line width=1.0pt, gray, dashed},
  lbl/.style={font=\scriptsize},
  blk/.style={font=\scriptsize\bfseries},
]
  \node[blk, anchor=east] at (-0.4, -1.5) {Block $A$};
  \foreach \m/\y in {0/0, 1/-1, 2/-2, 3/-3, 4/-4.2, 5/-5.2} {
    \draw[mode] (0,\y) -- (6.6,\y);
    \node[lbl, anchor=east] at (-0.05,\y) {$\m$};
  }
  \foreach \m/\y in {6/-6.4, 7/-7.4} {
    \draw[idle] (0,\y) -- (6.6,\y);
    \node[lbl, anchor=east, gray!65] at (-0.05,\y) {$\m$};
  }
  \node[blk, anchor=east] at (-0.4, -10.5) {Block $B$};
  \foreach \m/\y in {8/-9.0, 9/-10.0, 10/-11.0, 11/-12.0, 12/-13.2, 13/-14.2} {
    \draw[mode] (0,\y) -- (6.6,\y);
    \node[lbl, anchor=east] at (-0.05,\y) {$\m$};
  }
  \foreach \m/\y in {14/-15.4, 15/-16.4} {
    \draw[idle] (0,\y) -- (6.6,\y);
    \node[lbl, anchor=east, gray!65] at (-0.05,\y) {$\m$};
  }
  \draw[decorate,decoration={brace,amplitude=3pt}]
    (6.75,0.25) -- (6.75,-1.25)
    node[midway, right=3pt, lbl] {$0_A$: $|0\rangle_L$};
  \draw[decorate,decoration={brace,amplitude=3pt}]
    (6.75,-1.75) -- (6.75,-3.25)
    node[midway, right=3pt, lbl] {$1_A$: $|1\rangle_L$};
  \draw[decorate,decoration={brace,amplitude=3pt}]
    (6.75,-8.75) -- (6.75,-10.25)
    node[midway, right=3pt, lbl] {$0_B$: $|0\rangle_L$};
  \draw[decorate,decoration={brace,amplitude=3pt}]
    (6.75,-10.75) -- (6.75,-12.25)
    node[midway, right=3pt, lbl] {$1_B$: $|1\rangle_L$};
  \node[lbl, anchor=west] at (6.7,-4.7) {dumps};
  \node[lbl, anchor=west, gray!65] at (6.7,-6.9) {idle};
  \node[lbl, anchor=west] at (6.7,-13.7) {dumps};
  \node[lbl, anchor=west, gray!65] at (6.7,-15.9) {idle};
  \node[lbl, anchor=east, blue!60!black] at (-1.05, 0)
    {single photon $\rightarrow$};
  \node[lbl, anchor=east, blue!60!black] at (-1.05, -9.0)
    {single photon $\rightarrow$};
  \draw[fill=blue!12, draw=blue!60!black, rounded corners=1pt]
    (0.5,0.45) rectangle (1.5,-3.45);
  \node[lbl, blue!60!black] at (1.0,-1.5) {$U_{\rm prep}^A$};
  \draw[fill=blue!12, draw=blue!60!black, rounded corners=1pt]
    (0.5,-8.55) rectangle (1.5,-12.45);
  \node[lbl, blue!60!black] at (1.0,-10.5) {$U_{\rm prep}^B$};
  \draw[bs] (2.9,-2) -- (3.3,-11);
  \draw[bs] (3.5,-3) -- (3.9,-12);
  \node[lbl, red!70!black, anchor=west] at (3.95,-6.9)
    {CZ ($T{=}\tfrac13$)};
  \draw[dump] (2.3,0) -- (2.5,-4.2);
  \draw[dump] (2.5,-1) -- (2.7,-5.2);
  \draw[dump] (2.3,-9.0) -- (2.5,-13.2);
  \draw[dump] (2.5,-10.0) -- (2.7,-14.2);
  \draw[fill=green!12, draw=green!50!black, rounded corners=1pt]
    (4.8,0.45) rectangle (5.8,-3.45);
  \node[lbl, green!50!black] at (5.3,-1.5) {$R(\theta_A)$};
  \draw[fill=green!12, draw=green!50!black, rounded corners=1pt]
    (4.8,-8.55) rectangle (5.8,-12.45);
  \node[lbl, green!50!black] at (5.3,-10.5) {$R(\theta_B)H_B$};
\end{tikzpicture}%
}
\caption{Two-block layout on the $24$-mode processor
(circuit-mode indices).  One single photon enters each block
(circuit modes $0$ and $8$); the block-local preparation unitaries
$U_{\rm prep}^{A},U_{\rm prep}^{B}$ create the logical state on two
neutral rails (Eq.~\eqref{eq:logical}), here $|{+}\rangle_L$ on each
block.  The postselected controlled-$Z$ couples
rail $1$ of each block pairwise with $T = 1/3$ beam splitters
(solid, modes $2\!\leftrightarrow\!10$ and $3\!\leftrightarrow\!11$),
with matching $T = 1/3$ couplings from rail $0$ to local dump modes
($4,5,12,13$; dashed) for amplitude balancing.  Circuit modes $6,7$
(block $A$) and $14,15$ (block $B$) are idle (dotted); device modes
$16$--$23$ are unused.  Block-local readout applies $R(\theta_A)$ on
block $A$ and $R(\theta_B)H_B$ on block $B$ (the folded-in logical
Hadamard $H_B$, a Hadamard on $\mathcal{Q}_L$, acting first, then the
analyzer rotation;
Eqs.~\eqref{eq:rotation}--\eqref{eq:hadamard}) before rail
extraction; events with exactly one click in the logical ports of
each block are kept.  Line style, not color alone, distinguishes
gate couplings (solid), dump couplings (dashed), and idle modes
(dotted).}
\label{fig:layout}
\end{figure}

Block $A$ occupies circuit modes $0$--$7$ and block $B$ circuit
modes $8$--$15$; one photon is injected into each block (circuit
modes $0$ and $8$), for a two-photon, $16$-mode configuration on
the $24$-mode device.  We write ``circuit modes'' rather than
``chip modes'' because the archived records use the user-specified
circuit indices; the provider's compiled physical mapping is not
returned through the cloud interface.  Although the logical basis of
Eq.~\eqref{eq:logical} spans only four modes per block, each qubit is
embedded in the full eight-mode block: within a block, circuit modes
$0$--$3$ carry the two logical rails, circuit modes $4$--$5$ serve as
amplitude-balancing dumps (see the gate description below), and the
remaining two modes are idle (circuit modes $6,7$ in block $A$ and
$14,15$ in block $B$); device modes $16$--$23$ are unused
(Fig.~\ref{fig:layout}).

Operationally, the parity constraint defines the neutral-rail basis,
but no zero-sum syndrome measurement is performed; the encoding
therefore operates by postselection.
Events are retained only when exactly one photon exits in the logical
ports of each block (readout below), so population leaking into the
dump, idle, or uniform-sum directions produces clicks outside the
logical ports and is discarded rather than corrected.  The
single-block selectivity of that syndrome
was characterized separately in Ref.~\cite{q3paper}.  Consequently
the logical-port postselection rejects only the restricted class of
leakage errors
that move population out of the logical ports under this readout;
errors that remain within $\mathcal{Q}_L$ (or that leave one photon
in a logical port while corrupting the amplitude within $\mathcal{N}$)
are undetected, and the postselection does not distinguish parity
leakage from ordinary postselected-gate failure: both are absorbed
into the discarded branch.  In the ideal target these outcomes
constitute the $8/9$ failure branch; in practice, the discarded set
also includes leakage and other nonideal outcomes.  The residual
fraction of accepted events originating from such leakage was not
separately resolved in these runs, and it is not bounded by the
accepted-fraction figures of Table~\ref{tab:persetting}: because the
discarded branch mixes gate failure, loss, and leakage without
separating them, that acceptance rate constrains none of the three
individually (Sec.~\ref{sec:limits}).  Thus the present encoding is
postselected rather than error-corrected; no error-detected logical
Bell pair is demonstrated.

\textit{Entangling gate.}  The blocks are coupled by a
postselected controlled-$Z$ (CZ) gate of the
Ralph--Langford--Bell--White (RLBW) type~\cite{ralph2002,obrien2003},
generalized from single modes to two-mode rails: the two sub-modes
of rail $1$ of block $A$ are coupled pairwise to the corresponding
sub-modes of rail $1$ of block $B$ by beam splitters of
transmissivity $T = 1/3$ (modes $2\leftrightarrow 10$ and
$3\leftrightarrow 11$), while rail $0$ of each block is coupled to
a local dump rail by identical $T = 1/3$ splitters for amplitude
balancing.  Because both logical rails carry the same
$(e_i - e_j)/\sqrt{2}$ internal pattern, the pairwise coupling
commutes with the in-rail structure and the two-photon
interference reproduces the single-mode gate amplitudes exactly
(the $|11\rangle_L$ amplitude acquires the two-photon sign flip
$T - \mathcal{R} = -1/3$, with reflectivity $\mathcal{R} = 1 - T = 2/3$;
Appendix~\ref{app:cz}).  The gate
succeeds when one photon exits in the logical
ports of each block, with ideal probability $1/9$.

\textit{State preparation, readout, and target factorization.}
Logical preparations and measurement-basis rotations are compiled
into the same $16\times16$ unitary as the gate.  Here
$U_{xy}^{\mathrm{target}}$ is the $16\times16$ unitary on the used
circuit modes $0$--$15$ only; the eight idle device modes $16$--$23$
are omitted rather than embedded in a $24$-mode identity, and it is
these $16\times16$ matrices whose content hashes appear in
Table~\ref{tab:hashes}, so those digests compare cleanly, on the same
$16$-mode support, to the parameters of Table~\ref{tab:device}.  For
each CHSH setting pair $(x,y)$ the target circuit factorizes as
\begin{equation}
\label{eq:factorization}
  U_{xy}^{\mathrm{target}} =
  \bigl(U^{A}_{\mathrm{read}}(x) \oplus U^{B}_{\mathrm{read}}(y)\bigr)\,
  U_{\mathrm{CZ}}\,
  \bigl(U^{A}_{\mathrm{prep}} \oplus U^{B}_{\mathrm{prep}}\bigr),
\end{equation}
where each block-local preparation unitary $U^{A/B}_{\mathrm{prep}}$
is an $8\times8$ mode transformation constrained to map the injected
physical input mode $e_0$ to the desired logical state.  For every
CHSH setting both blocks are prepared in
$|{+}\rangle_L = \tfrac{1}{\sqrt2}(|0\rangle_L + |1\rangle_L)
= \tfrac12(e_0 - e_1 + e_2 - e_3)$, so
$U^{A/B}_{\mathrm{prep}} e_0 = |{+}\rangle_L$ in block-local
coordinates (the globally injected block-$B$ mode is circuit mode
$8$); no particular logical action of this unitary on arbitrary inputs
in $\mathcal{Q}_L$ is assumed.  The
entangling transformation $U_{\mathrm{CZ}}$ is the rail-wise
postselected controlled-$Z$ above, and the block-local readout
unitaries $U^{A}_{\mathrm{read}}(x)$, $U^{B}_{\mathrm{read}}(y)$ act
on the logical subspace as the analyzer rotation $R(\theta)$ of
Eq.~\eqref{eq:rotation} (with the fixed logical Hadamard $H_B$ of
Eq.~\eqref{eq:hadamard} folded into the block-$B$ readout), followed
by rail extraction (the readout step that routes each logical rail to
a single output mode: in block-local coordinates
$(e_0-e_1)/\sqrt2 \mapsto e_0$ and $(e_2-e_3)/\sqrt2 \mapsto e_2$, with
block $B$ shifted by $+8$ and the orthogonal complements routed to
nonlogical ports), so that a threshold click in port $0$ ($2$) of
a block records logical outcome $0$ ($1$).  The direct sums act block-diagonally on circuit
modes $0$--$7$ (block $A$) and $8$--$15$ (block $B$), with no
cross-block terms.  By construction the four targets share an
identical preparation (both blocks in $|{+}\rangle_L$) and the
identical postselected controlled-$Z$, and differ only in the two
block-diagonal analyzer factors; the preparation and readout carry no
cross-block terms, and the one-photon-per-logical-block success
projector is setting-independent.  Equation~\eqref{eq:factorization}
holds exactly for each target: the circuit-construction code
in the reproducibility package assembles the four
$16\times16$ matrices as the product
$U_{\mathrm{read}}\,U_{\mathrm{CZ}}\,U_{\mathrm{prep}}$ and verifies
unitarity numerically (residual
$\max|U^\dagger U - \mathbb{1}| < 10^{-10}$).
These statements apply to the submitted target matrices.  Each target
is transpiled to the physical interferometer server-side by the cloud
compiler; the compiled mapping is not returned through the interface,
and the executed transformation is assumed to preserve the intended
common-preparation, local-analyzer structure of the targets.  The
archive identifies every submitted circuit by its cloud job identifier
(for cross-checking against the provider's job console), not by a
returned compiled matrix.

Events are postselected on exactly one click among the logical
ports of block $A$ and one among those of block $B$, with no
clicks elsewhere; all other outcomes (dumps, nonlogical-port
leakage, bunching) are discarded.  In the ideal target these constitute
the $8/9$ failure branch; in practice the discarded set also includes
leakage and other nonideal outcomes.  In the ideal single-pair model,
writing $\Pi_{1,1}$ for the projector onto one photon in each block
and $\Pi_L$ for the projector onto
$\mathcal{Q}_L\otimes\mathcal{Q}_L$, and with $\rho$ the
pre-postselection two-photon output density operator, the accepted
(postselected) state is
\begin{equation}
\label{eq:rhops}
  \rho_{\mathrm{ps}} =
  \frac{\Pi_L\,\Pi_{1,1}\,\rho\,\Pi_{1,1}\,\Pi_L}
       {\operatorname{Tr}\!\bigl(\Pi_L\,\Pi_{1,1}\,\rho\,\Pi_{1,1}\,\Pi_L\bigr)} .
\end{equation}
Equation~\eqref{eq:rhops} is an idealized single-pair description; it
is subject to the threshold-detector caveat stated immediately below
and quantified in Sec.~\ref{sec:limits}, because the physical
detectors are non-number-resolving.
Because acceptance requires exactly one photon in each block in this
model, the one-photon sector of block $A$ tensored with that of
block $B$ supplies the effective bipartite tensor-product structure
$\mathcal{Q}_L^{A}\otimes\mathcal{Q}_L^{B}$ on which the block-local
analyzer rotations act as local observables.  The actual hardware
uses threshold detectors, so number-degenerate multiphoton events
that survive the click filter are not literally described by
Eq.~\eqref{eq:rhops}; they are treated as a residual contrast-loss
channel in Sec.~\ref{sec:limits}.

\textit{CHSH settings and measurements.}  We assign the readout
outcomes $0 \mapsto +1$ and $1 \mapsto -1$, so that each block-local
measurement is dichotomic.  A logical analyzer at angle $\theta$ is
the real-valued rotation
\begin{multline}
\label{eq:rotation}
  R(\theta) =
  \begin{pmatrix} \cos\theta & -\sin\theta \\
                  \sin\theta & \phantom{-}\cos\theta \end{pmatrix},
  \\
  R^\dagger(\theta)\,Z\,R(\theta) = \cos(2\theta)\,Z - \sin(2\theta)\,X,
\end{multline}
so measuring $Z$ after $R(\theta)$ realizes the observable
$\cos(2\theta)\,Z - \sin(2\theta)\,X$ on $\mathcal{Q}_L$.  Applied to
the equal-superposition inputs
$|{+}\rangle_L = \tfrac{1}{\sqrt2}(|0\rangle_L + |1\rangle_L)$ on
both blocks (with
$|{-}\rangle_L = \tfrac{1}{\sqrt2}(|0\rangle_L - |1\rangle_L)$), the
controlled-$Z$ gives
$\mathrm{CZ}\,|{+}{+}\rangle_L = \tfrac{1}{\sqrt2}(|0{+}\rangle_L +
|1{-}\rangle_L)$, and a local Hadamard $H_B$ on block $B$ maps this to
the maximally entangled Bell state,
\begin{equation}
\label{eq:hadamard}
  (I \otimes H_B)\,\mathrm{CZ}\,|{+}{+}\rangle_L
  = |\Phi^+\rangle
  = \tfrac{1}{\sqrt2}\bigl(|00\rangle_L + |11\rangle_L\bigr).
\end{equation}
We fold $H_B$ into the block-$B$ readout, so that block $B$ applies
$R(\theta_B)H_B$ (read right to left: $H_B$ acts first, then the
analyzer rotation $R(\theta_B)$).  We use the standard CHSH
settings $(a_0, a_1) = (0, \pi/4)$ for block $A$ and
$(b_0, b_1) = (\pi/8, -\pi/8)$ for block $B$, applied as the real
rotations $R(\theta)$ of Eq.~\eqref{eq:rotation} before rail
extraction.

\textit{Returned-sample probability space.}  We fix the sample space
on which every count and score below is defined.  The remote processor
draws $N_{\rm clk}$ raw source-clock trials; these are not
available through the interface and enter no denominator here.  The
provider applies its \texttt{min\_detected\_photons}${}=2$
returned-sample filter and delivers only the passing threshold
patterns, of which each job returns a fixed number
$N_{\rm ret}=5{,}000$ two-photon patterns.  This returned set is the
denominator of every rate we quote.  We then apply the offline logical
postselection predicate $\Lambda$ (exactly one click in the two
logical ports of each block, none elsewhere), leaving
$N_{\rm acc}=\sum_{ij}N_{ij}$ accepted logical coincidences.  Thus
\begin{equation}
\label{eq:samplespace}
  N_{\rm clk}\ \xrightarrow{\ \texttt{min\_detected\_photons}\ }\
  N_{\rm ret}\ \xrightarrow{\ \Lambda\ }\ N_{\rm acc}
  = \sum_{ij} N_{ij},
\end{equation}
and every correlator, marginal, and score reported here is
conditional on the provider filter: it is a statement about the
returned-then-accepted set, not about the inaccessible source clock.
For each setting pair we form the correlator
\begin{equation}
\label{eq:E}
  E(a,b) = \frac{N_{00} + N_{11} - N_{01} - N_{10}}
                {N_{00} + N_{11} + N_{01} + N_{10}},
\end{equation}
where $N_{ij}$ counts accepted logical coincidences (elements of
$N_{\rm acc}$) with logical outcome $i$ on block $A$ and $j$ on
block $B$, and combine the four pairs into
\begin{equation}
\label{eq:S}
  S = E(a_0,b_0) + E(a_0,b_1) + E(a_1,b_0) - E(a_1,b_1).
\end{equation}
For $|\Phi^+\rangle$ the ideal correlators follow from
Eq.~\eqref{eq:rotation} as
\begin{equation}
\label{eq:Ecos}
  E(\theta_A,\theta_B) = \cos\bigl[2(\theta_A - \theta_B)\bigr],
\end{equation}
so the chosen angles give $E = +1/\sqrt2$ for the first three pairs
and $E = -1/\sqrt2$ for $(a_1,b_1)$; all four terms of
Eq.~\eqref{eq:S} then contribute $+1/\sqrt2$ and $S = 2\sqrt2$, the
ideal submitted-target value.  For any local dichotomic assignments
$A_x, B_y \in \{\pm1\}$ the identity
$A_0(B_0+B_1) + A_1(B_0-B_1) = \pm2$ holds deterministically (one of
$B_0 \pm B_1$ vanishes), so $|S| \le 2$; averaging over any product
distribution and, by convexity, over any separable state preserves
the bound~\cite{chsh1969}.  Accordingly, interpreting $|S|>2$ as an
entanglement witness requires the common-state, local-analyzer,
setting-independent-postselection, and fair-sampling premises examined
in Secs.~\ref{sec:results}--\ref{sec:limits}.

\textit{Estimator symbols.}  We use a fixed notation throughout:
$S$ is the operational score on accepted counts; $S_i$ a single-pass
score; $S_{\rm count}$ the naive count-pooled score;
$S_{\rm pass}$ the pass-level (random-effects) pooled estimate;
$S^{\rm rw}$ a fixed-ratio efficiency-reweighted per-pass score; and
$S^{\rm rw}_{\rm count}$ (respectively $S^{\rm rw}_{\rm pass}$) its
count-pooled (respectively pass-level) counterpart.  Reweighted scores
are archived-metadata model scenarios, not corrected platform scores.

\section{Experimental protocol}
\label{sec:protocol}

All data were acquired on the 24-mode Quandela Belenos
processor~\cite{quandela_belenos,quandela_catalog}
(\texttt{qpu:belenos}) through the Perceval
interface~\cite{perceval}, as standard external cloud jobs;
state preparation, the entangling gate, and readout are compiled
to the native interferometer mesh under the standard universal
linear-optical
model~\cite{klm2001,reck1994,clements2016,carolan2015,flamini2019}.
Photons are generated by the platform's quantum-dot single-photon
source: a semiconductor quantum dot in an optical microcavity that,
under pulsed excitation, emits near-transform-limited single photons
on demand.  Simultaneously maintaining high single-photon purity and
indistinguishability requires active stabilization; in our modeling, partial distinguishability and
phase drift are treated jointly as the dominant contrast-loss
contribution to the postselected gate reported here
(Sec.~\ref{sec:limits}), with multiphoton emission a much
smaller correction.  A single emitter feeds a temporal-to-spatial
demultiplexer that routes successive pulses into the two block
inputs.  This source's $4.94$~MHz single-photon clock, second-order
correlation $g^{(2)}(0)$, and Hong--Ou--Mandel (HOM)
visibility~\cite{hong1987} on this processor were characterized in
Ref.~\cite{q3paper}.  Two CHSH sessions were
run on 4~July~2026, separated by a same-day platform recalibration:
the first at platform-reported HOM visibility $92.0\%$ at session
start ($90.3\%$ from its third pass onward), the second at $91.5\%$
throughout; both at second-order correlation $g^{(2)}(0) = 0.011$ and
clock $4.94$~MHz (Table~\ref{tab:device} collects the device and
operating parameters).  The cloud endpoint was operated with
threshold-detection semantics (non-number-resolving at the returned-click
level); postselection on one click per block discards multi-click events,
but number-degenerate multiphoton events that produce a single
threshold click remain part of the residual-multiphoton systematic.
Residual multiphoton emission is treated as a modeled contrast-loss
channel rather than a sign-definite guarantee: with source
$g^{(2)}(0) = 0.011$ and transmittance $\eta \approx 5\%$, the
dominant path (a two-photon emission in one block with one photon
lost before detection) leaves a ``lost-twin'' survivor at the
percent level per block.  Under the assumption that such survivors
are uncorrelated between the two blocks they dilute the correlators;
we estimate this channel for the present two-photon configuration in
Sec.~\ref{sec:limits}, instead of
relying on the single-block figures of Ref.~\cite{q3paper}.

\begin{table}[tbp]
\caption{Device and operating parameters for the CHSH sessions}
\label{tab:device}
\begin{tabularx}{\columnwidth}{@{}l>{\raggedright\arraybackslash}X@{}}
\toprule
Parameter & Value \\
\midrule
Processor            & Belenos (MosaiQ-12) \\
Cloud endpoint       & \texttt{qpu:belenos} \\
Native model         & universal linear optics \\
Spatial modes        & $24$ \\
Photon capacity      & $12$ \\
Photon source        & quantum dot \\
Source clock         & $4.94$~MHz \\
HOM, S1\,/\,S2 (vendor/dashboard) & $92.0\%\,/\,91.5\%$ \\
$g^{(2)}(0)$ (vendor/dashboard) & $0.011$ \\
Transmittance $\eta$ & ${\approx}\,5\%$ \\
Detectors            & threshold semantics (non-PNR at returned-click level) \\
\midrule
Modes used           & $16$ (two $8$-mode blocks) \\
Photons used         & $2$ \\
Entangling gate      & postselected CZ, $p{=}1/9$ \\
Returned samples/job & $5000$ (\texttt{max\_samples}) \\
\bottomrule
\end{tabularx}
\begin{minipage}{\linewidth}\vspace{4pt}\footnotesize\noindent\emph{Notes.}~
Platform-reported calibrations at session start for the two
CHSH sessions (S1/S2) of 4~July~2026, submitted through the Perceval
cloud interface~\cite{perceval}.  HOM figures are dashboard
values at pass start, not contemporaneous in-job measurements;
the reported HOM value was $90.3\%$ from the third pass onward, and
the per-photon transmittance is $4.9$--$5.2\%$.  ``Returned
samples'' are Perceval samples of interest (\texttt{max\_samples}),
not source-clock trials (see text); every archived CHSH job returned
exactly $5{,}000$ such samples.  The quoted $p{=}1/9$ is the
ideal target gate-success probability, not a measured acceptance
rate.  Gate-characterization jobs used
$20{,}000$ returned samples except one $5{,}000$-sample pilot, which
enters no reported quantity.  Detector row: threshold-detection
semantics at the returned-click level under the cloud interface, not a
separate physical-detector certification beyond vendor documentation.
\end{minipage}
\end{table}

\begin{table}[tbp]
\caption{Run parameters and offline postselection rule}
\label{tab:runparams}
\begin{tabularx}{\columnwidth}{@{}l>{\raggedright\arraybackslash}X@{}}
\toprule
Parameter & Value \\
\midrule
Sampler \texttt{max\_samples}          & $5000$ \\
Returned-sample filter                 & \texttt{min\_detected\_photons}${}=2$ \\
Perceval / \texttt{perceval-quandela}  & $1.2.4$ \\
NumPy (analysis environment)           & $2.4.4$ \\
Offline postselection                  & one logical-port click per block \\
\quad block-$A$ logical ports          & circuit modes $0,2$ \\
\quad block-$B$ logical ports          & circuit modes $8,10$ \\
Predicate SHA-256 (first 16 hex)       & \texttt{6bae83187a6bba37} \\
\bottomrule
\end{tabularx}
\begin{minipage}{\linewidth}\vspace{4pt}\footnotesize\noindent\emph{Notes.}~
Summary of the acquisition and analysis parameters used
for every archived CHSH job.  \texttt{max\_samples} is the requested
number of returned two-photon samples per job (Sec.~\ref{sec:protocol});
\texttt{min\_detected\_photons}${}=2$ is the provider-side returned-sample
filter applied before delivery.  The pinned software versions are those
of the reproducibility package \texttt{requirements.txt}.  The offline
postselection accepts an event iff exactly one click falls in the two
logical ports of each block and no click elsewhere; its full source
(\texttt{classify}, together with the port lists \texttt{PORT\_A} and
\texttt{PORT\_B}) is digested by \texttt{verify\_targets.py} to the
SHA-256 predicate whose first $16$ hex digits are
\texttt{6bae83187a6bba37} (Ref.~\cite{phaseb_zenodo}).
\end{minipage}
\end{table}

\begin{table}[tbp]
\caption{Port-to-outcome map and readout-efficiency source}
\label{tab:portmap}
\centering\footnotesize
\setlength{\tabcolsep}{4pt}
\begin{tabular}{@{}lccc@{}}
\toprule
Logical outcome & Circuit mode & S1 $t_{\rm out}$ & S2 $t_{\rm out}$ \\
\midrule
$A=0$ & $0$  & $119.99$ & $104.98$ \\
$A=1$ & $2$  & $78.75$  & $66.15$  \\
$B=0$ & $8$  & $117.60$ & $117.82$ \\
$B=1$ & $10$ & $118.27$ & $121.68$ \\
\midrule
$\kappa_A = t_{\rm out}(A{=}0)/t_{\rm out}(A{=}1)$
      & --- & $1.524$ & $1.587$ \\
$\kappa_B = t_{\rm out}(B{=}0)/t_{\rm out}(B{=}1)$
      & --- & $0.994$ & $0.968$ \\
\bottomrule
\end{tabular}
\begin{minipage}{\linewidth}\vspace{4pt}\footnotesize\noindent\emph{Notes.}~
Mapping from circuit output modes to logical outcomes and the
start-of-session port transmittances $t_{\rm out}$ from which the
readout-imbalance ratios $\kappa_A,\kappa_B$ of Table~\ref{tab:chsh}
and Appendix~\ref{app:readout} are formed.  Modes $0\!\to\!A{=}0$,
$2\!\to\!A{=}1$, $8\!\to\!B{=}0$, $10\!\to\!B{=}1$; the block-$A$ ratio
is the dominant imbalance reweighted in the sensitivity analysis.
Precise meaning of $t_{\rm out}$: these are platform-reported
per-mode relative output transmittances (arbitrary units) read from
the cloud dashboard at session start, indexed by
user-facing circuit mode, with no post-compilation
re-measurement and no provider uncertainties; whether they refer to
physical chip outputs or to a provider-level virtual-port calibration
is not documented by the interface and is therefore not auditable.
The values were recorded at session start and not re-measured within a
session.
\end{minipage}
\end{table}

Because the interference contrast of a postselected gate is
degraded both by photon distinguishability and by slow phase drift
of the on-chip interferometer mesh that realizes the compiled
unitary (the two-photon Hong--Ou--Mandel interference occurs within
this same mesh; there is no separate interferometer) over the
acquisition window, the protocol was designed around short
acquisitions: each job requested $5{,}000$
returned samples of interest (completing in minutes on the
quantum processing unit (QPU) clock), and the four CHSH settings were
submitted back-to-back as one
pass, yielding a complete estimate in a short acquisition
window.  Each session comprised four consecutive passes ($16$ jobs),
providing four separate complete estimates of $S$ plus drift
information between passes; the second session repeated the protocol
end-to-end after a same-day recalibration.  The passes were
sequential on one day and shared hardware, source, compiler, and
calibration history, so we do not assert their statistical
independence; the between-pass structure is analyzed in
Sec.~\ref{sec:results}.

\textit{Sampling semantics.}  In Perceval a job requests a target
number of returned samples of interest (the \texttt{Sampler}
\texttt{max\_samples} parameter), which the remote processor
produces by drawing physical source-clock shots, bounded by
\texttt{max\_shots} and \texttt{max\_shots\_per\_call}, and
returning only detector patterns that pass a physical
detected-photon filter (\texttt{min\_detected\_photons})~%
\cite{perceval,perceval_sampler}.  A ``$5{,}000$-sample job'' thus
requests $\texttt{max\_samples} = 5{,}000$ returned two-photon
samples of interest, not $5{,}000$ source-clock trials: at
the $4.94$~MHz clock, ${\approx}5\%$ per-photon transmission, and
the ${\sim}1/9$ postselected-gate branch, a naive $5{,}000$ source
trials would yield only of order
$5000\times(0.05)^2\times(1/9)\sim 1$ accepted coincidence, whereas
each archived CHSH job returned exactly $5{,}000$ two-photon samples
and ${\sim}700$--$900$ accepted logical coincidences after offline
postselection.  The denominator of every rate we quote is the
number of returned two-photon samples, and logical postselection
(one click in the logical ports of each block) is applied to those
returned patterns.  For every job the reproducibility package
archives the cloud job identifier, setting label, pass index,
timestamp, recycling flag, platform HOM value where recorded,
requested returned-sample count, and the full returned click-pattern
counts used for offline logical filtering.  The analysis scripts
define the logical postselection expression, detector convention, and
threshold filter used to regenerate the reported tables; lower-level
provider shot-limit and performance metadata were not returned
uniformly and are not required for the offline verification.

\textit{Acquisition order and provenance.}  Within every pass the
four settings were acquired sequentially in the fixed order
$(a_0,b_0)$, $(a_0,b_1)$, $(a_1,b_0)$, $(a_1,b_1)$.  Because the order
was not randomized or counterbalanced, setting identity and
acquisition time are confounded; the consequences are analyzed in
Sec.~\ref{sec:results}.  Each session ran a fixed four-pass
protocol under an author-set live-HOM acceptance gate in the
acquisition script: before each session we set a nominal target
dashboard HOM of $90.5\%$ (the script default), with a hard abort
threshold at $90.0\%$ ($0.5$ percentage points below the nominal
target), so the run aborted only if the dashboard value fell below
$90.0\%$.  This criterion was fixed before the CHSH passes; the
lowest recorded pass HOM was $90.3\%$ (session~1), which was accepted
under the $90.0\%$ hard abort threshold (below the $90.5\%$ nominal
target), and no CHSH job was aborted by it.  These HOM figures, like the other platform
parameters of Table~\ref{tab:device}, are dashboard values read at pass
start rather than measurements made within the jobs themselves.  No
completed pass was discarded and no CHSH job in
either session failed, was cancelled, or was excluded.  The archived
manifest lists all $59$ jobs underlying the study: $18$ session-1 CHSH
jobs, $16$ session-2 CHSH jobs, and $25$ gate-characterization jobs,
each with its cloud job identifier, sample
count, and pass-start HOM value~\cite{phaseb_zenodo}.  The complete
manifest, including full job identifiers, is supplied as
\texttt{manifest/job\_manifest.csv} in the archived data package, so
that every reported quantity can be traced to an identified job.
Per-job wall-clock
timestamps are archived for the $34$ CHSH-session jobs; the $25$
gate-characterization records carry no timestamp field, and raw
source-clock shot counts are not available from the interface for any
job (Sec.~\ref{sec:protocol}).  We refer to the resulting data set
throughout as the fixed-order reference acquisition.

The platform's built-in photon-recycling mitigation was disabled for
all CHSH jobs, since event reconstruction alters postselected
two-photon statistics in a way that is difficult to audit; a two-job
A/B comparison (one CHSH setting acquired with recycling on and off)
is summarized in Sec.~\ref{sec:results} and archived in full in the
reproducibility package~\cite{phaseb_zenodo}.  A separate set of long-window
($20{,}000$-sample) acquisitions on the same submitted target circuits was
exploratory and informed the short-window protocol; the
gate-characterization truth-table and phase checks
(Sec.~\ref{sec:results}) are drawn from that set.  Those long-window
correlator acquisitions were not included in the primary estimate
because they preceded the final acquisition protocol; all such
acquisitions, including failed or repeated jobs, are reported in the
reproducibility package~\cite{phaseb_zenodo}.

\section{Results}
\label{sec:results}

\begin{table}[tbp]
\caption{Conditional controlled-$Z$ diagnostic results}
\label{tab:tt}
\centering\small
\setlength{\tabcolsep}{4pt}
\begin{tabular}{@{}lrcc@{}}
\toprule
Configuration & $n$ & frac. & $P(\mathrm{correct})$ (95\% CI) \\
\midrule
$|00\rangle_L$ truth table & $10763$ & $0.269$ & $0.970\,[0.966,0.973]$ \\
$|01\rangle_L$ truth table & $7309$  & $0.183$ & $0.959\,[0.954,0.963]$ \\
$|10\rangle_L$ truth table & $6499$  & $0.162$ & $0.929\,[0.922,0.935]$ \\
$|11\rangle_L$ truth table & $5849$  & $0.146$ & $0.981\,[0.977,0.984]$ \\
$|{+}0\rangle_L$ phase check & $7432$ & $0.186$ & $0.959\,[0.955,0.964]$ \\
$|{+}1\rangle_L$ phase check & $5166$ & $0.129$ & $0.803\,[0.792,0.813]$ \\
\midrule
Total & $43018$ & $0.179$ & --- \\
\bottomrule
\end{tabular}
\begin{minipage}{\linewidth}\vspace{4pt}\footnotesize\noindent\emph{Notes.}~
Conditional computational-basis label preservation
(truth-table rows) and phase-flip diagnostics (phase-check rows) for
the controlled-$Z$, from the exploratory long-window
gate-characterization set.  $P(\mathrm{correct})$ is the fraction of
accepted logical coincidences returning the ideal outcome,
conditional on acceptance; uncertainties are $95\%$ Wilson score
intervals~\cite{wilson1927} on $n$ coincidences.  These are
diagnostic correctness probabilities, not a gate process
fidelity.  Each configuration was acquired as two
$20{,}000$-sample jobs, so ``frac.''\ is $n/40{,}000$ returned
two-photon samples, on the same denominator convention as
Table~\ref{tab:persetting}; the acceptance rates of this exploratory
set ($0.13$--$0.27$) span a wider range than those of the CHSH passes
($0.147$--$0.179$) because the configurations differ in logical input
rather than analyzer setting.  For the two phase-check rows, block~$A$
is read in the $X$ basis and the outcome counted as correct is the
$X$-basis eigenstate the ideal controlled-$Z$ produces on block~$A$:
$|{+}\rangle_A$ ($X{=}{+}1$, readout outcome $0$) for the
$|{+}0\rangle_L$ row (no conditional flip), and $|{-}\rangle_A$
($X{=}{-}1$, readout outcome $1$) for the $|{+}1\rangle_L$ row,
reflecting the conditional sign flip
$\mathrm{CZ}\,|{+}1\rangle=|{-}1\rangle$.
\end{minipage}
\end{table}

\textit{Gate diagnostics.}  The controlled-$Z$ transformation
preserves all computational-basis labels; its minus sign on
$|11\rangle_L$ is not observable in a $Z$-basis truth-table
measurement, which cannot distinguish controlled-$Z$ from the
identity.  Conditional on acceptance, the four computational-basis
inputs returned the correct label with probability $0.93$--$0.98$
(Table~\ref{tab:tt}).  The phase checks prepared $|{+}0\rangle_L$ and
$|{+}1\rangle_L$ and read block $A$ in the $X$ basis, probing the
conditional sign flip $\mathrm{CZ}|{+}1\rangle = |{-}1\rangle$; the
correct outcome dominated in both cases ($0.959$ and $0.803$; the
$X$-basis check is interference-sensitive and correspondingly more
exposed to drift).  The rail-asymmetric $|{+}1\rangle_L$ shortfall
sits in the same direction as the block-$A$ transmittance imbalance
analyzed below.  The same
coherent two-photon interference that fixes the CHSH correlators
governs this $X$-basis check, so the lower $|{+}1\rangle_L$ value is
consistent with the shortfall of the measured $S$ below $2\sqrt2$ and
motivates the short-pass acquisition (Sec.~\ref{sec:protocol}) used
for the CHSH measurement.  These measurements are descriptive gate
diagnostics rather than process tomography, and the exploratory
long-window data are excluded from the primary CHSH corpus.  Because
the effective contrast $v$ is inferred from the CHSH data themselves,
the agreement is qualitative rather than an independent validation.

\begin{table}[tbp]
\caption{CHSH estimates by acquisition pass}
\label{tab:chsh}
\centering\footnotesize
\setlength{\tabcolsep}{2.5pt}
\begin{tabular}{@{}llcccc@{}}
\toprule
Sess. & Pass & HOM & $n$ & $S$ & $S^{\rm rw}$ \\
\midrule
1 & 0 & 92.0 & 3447 & $2.202 \pm 0.057$ & $2.040 \pm 0.062$ \\
1 & 1 & 92.0 & 3390 & $2.371 \pm 0.056$ & $2.214 \pm 0.060$ \\
1 & 2 & 90.3 & 3332 & $2.541 \pm 0.053$ & $2.452 \pm 0.057$ \\
1 & 3 & 90.3 & 3382 & $2.485 \pm 0.054$ & $2.385 \pm 0.058$ \\
2 & 0 & 91.5 & 3307 & $2.618 \pm 0.053$ & $2.531 \pm 0.057$ \\
2 & 1 & 91.5 & 3316 & $2.551 \pm 0.053$ & $2.472 \pm 0.057$ \\
2 & 2 & 91.5 & 3273 & $2.552 \pm 0.054$ & $2.452 \pm 0.058$ \\
2 & 3 & 91.5 & 3386 & $2.579 \pm 0.053$ & $2.509 \pm 0.056$ \\
\midrule
1 & pooled & --- & 13551 & $2.397 \pm 0.028$ & $2.269 \pm 0.030$ \\
2 & pooled & --- & 13282 & $2.575 \pm 0.027$ & $2.491 \pm 0.029$ \\
\multicolumn{2}{@{}l}{Combined} & --- & 26833 & $2.485 \pm 0.019$
  & $2.380 \pm 0.021$ \\
\bottomrule
\end{tabular}
\begin{minipage}{\linewidth}\vspace{4pt}\footnotesize\noindent\emph{Notes.}~
Each pass is a complete four-setting estimate from four
consecutive $5{,}000$-sample jobs; $n$ is the total number of accepted
logical coincidences.  All quoted uncertainties are conditional
shot-noise errors: raw-$S$ uncertainties use binomial propagation
[Eq.~\eqref{eq:sigmaE}], whereas reweighted uncertainties use the
multinomial delta method (Appendix~\ref{app:readout}).  HOM is the
platform-reported visibility at pass start.  $S^{\rm rw}$ reweights
the block-$A$ outcome-$1$ counts by the session-specific archived
start-of-session port-transmittance ratios
$\kappa_{A,1}=1.524$ and $\kappa_{A,2}=1.587$
(Appendix~\ref{app:readout}); the corresponding block-$B$ ratios are
$0.994$ (session~1) and $0.968$ (session~2).  The reweighting is a
fixed-ratio sensitivity descriptor, not a contemporaneously calibrated
correction.  The passes show excess dispersion in session~1
($Q/\nu = 7.2$; unmodeled between-pass variation of unidentified
origin) but not in session~2 ($Q/\nu = 0.36$).  The pass-level
sensitivity analyses (Sec.~\ref{sec:results}) use the pass as the
pass-level analysis unit rather than these count-pooled shot-noise errors.
\end{minipage}
\end{table}

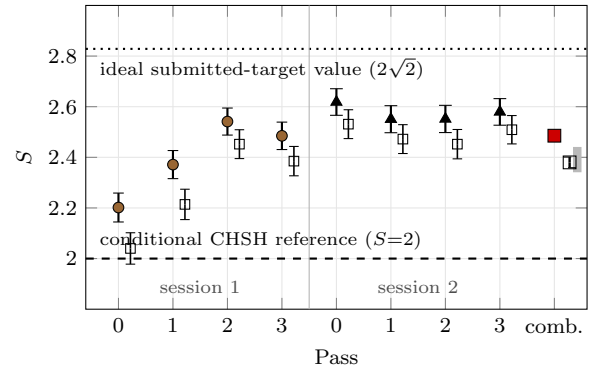
\begin{figure}[tbp]
\centering
\begin{tikzpicture}
\begin{axis}[
  width=0.95\columnwidth, height=5.6cm,
  xlabel={Pass}, ylabel={$S$},
  xmin=-0.6, xmax=8.6,
  ymin=1.8, ymax=3.0,
  xtick={0,1,2,3,4,5,6,7,8},
  xticklabels={0,1,2,3,0,1,2,3,comb.},
  ytick={2.0,2.2,2.4,2.6,2.8},
  grid=major, grid style={gray!20},
  tick label style={font=\footnotesize},
  label style={font=\footnotesize},
]
  \addplot[black, dashed, thick, domain=-0.6:8.6] {2};
  \addplot[black, dotted, thick, domain=-0.6:8.6] {2.8284};
  \node[font=\scriptsize, black, anchor=south west]
    at (axis cs:-0.5,2.0) {conditional CHSH reference ($S{=}2$)};
  \node[font=\scriptsize, black, anchor=north west]
    at (axis cs:-0.5,2.828) {ideal submitted-target value ($2\sqrt{2}$)};
  \draw[gray!60] (axis cs:3.5,1.8) -- (axis cs:3.5,3.0);
  \node[font=\scriptsize, gray!60!black, anchor=south]
    at (axis cs:1.5,1.83) {session 1};
  \node[font=\scriptsize, gray!60!black, anchor=south]
    at (axis cs:5.5,1.83) {session 2};
  \addplot+[only marks, mark=*, mark size=2pt, black,
    error bars/.cd, y dir=both, y explicit,
    error bar style={line width=0.8pt}]
  coordinates {
    (0, 2.2017) +- (0, 0.0571)
    (1, 2.3713) +- (0, 0.0555)
    (2, 2.5413) +- (0, 0.0534)
    (3, 2.4849) +- (0, 0.0542)
  };
  \addplot+[only marks, mark=triangle*, mark size=2.4pt, black,
    error bars/.cd, y dir=both, y explicit,
    error bar style={line width=0.8pt}]
  coordinates {
    (4, 2.6185) +- (0, 0.0525)
    (5, 2.5506) +- (0, 0.0534)
    (6, 2.5516) +- (0, 0.0538)
    (7, 2.5794) +- (0, 0.0525)
  };
  \addplot+[only marks, mark=square, mark size=1.8pt, black,
    error bars/.cd, y dir=both, y explicit,
    error bar style={line width=0.6pt, black}]
  coordinates {
    (0.22, 2.040) +- (0, 0.062)
    (1.22, 2.214) +- (0, 0.060)
    (2.22, 2.452) +- (0, 0.057)
    (3.22, 2.385) +- (0, 0.058)
    (4.22, 2.531) +- (0, 0.057)
    (5.22, 2.472) +- (0, 0.057)
    (6.22, 2.452) +- (0, 0.058)
    (7.22, 2.509) +- (0, 0.056)
  };
  \addplot+[only marks, mark=square*, mark size=2.4pt,
    black,
    error bars/.cd, y dir=both, y explicit,
    error bar style={line width=0.9pt}]
  coordinates { (8, 2.4854) +- (0, 0.0192) };
  \draw[line width=3.2pt, gray!55]
    (axis cs:8.42,2.341) -- (axis cs:8.42,2.441);
  \addplot+[only marks, mark=square, mark size=2.4pt,
    black,
    error bars/.cd, y dir=both, y explicit,
    error bar style={line width=0.9pt}]
  coordinates { (8.28, 2.380) +- (0, 0.021) };
\end{axis}
\end{tikzpicture}
\caption{CHSH parameter $S$ for the eight passes of the two sessions
(filled circles: session~1 raw; filled triangles: session~2 raw, after a
same-day recalibration) and the count-pooled combined results.  Open
squares (offset in $x$ for clarity) are the session-specific
fixed-ratio reweighted per-pass values $S^{\rm rw}$
($\kappa_{A,1}=1.524$, $\kappa_{A,2}=1.587$;
Table~\ref{tab:chsh}); at ``comb.'' the filled square is the raw
count-pooled $S_{\rm count} = 2.485$ and the open square the
reweighted count-pooled $S^{\rm rw}_{\rm count} = 2.380$.
Thin bars are conditional (shot-noise) $\pm1\sigma$ errors; the
between-pass spread is wider (Sec.~\ref{sec:results}).  The thick
gray bar at the combined reweighted point is an \emph{ad hoc}
$\kappa$ stress scan spanning $2.341$--$2.441$ (equal-session ratios
$\kappa\in[1.2,1.8]$; Eq.~\eqref{eq:kappa_band_wide}); it is a model
scan, not a statistical interval, and is distinct from the thin
shot-noise whiskers.  The dashed $S=2$ line is shown as a
conditional CHSH reference; its entanglement-witness interpretation
requires the assumptions discussed in Sec.~\ref{sec:limits}.  The
ideal submitted-target value $2\sqrt{2}$ is shown dotted.}
\label{fig:chsh}
\end{figure}

\textit{Pass-level CHSH results.}  The primary analysis uses the pass as
the unit of analysis and reports per-pass scores before any pooled figure.
Table~\ref{tab:chsh} and Fig.~\ref{fig:chsh} report the eight passes of
the two sessions.  The operational per-pass score $S$, defined on the
accepted logical coincidences [Eqs.~\eqref{eq:E}--\eqref{eq:S}], has
session means $2.40$ (session~1) and $2.58$ (session~2) with sample
standard deviations $0.15$ and $0.03$.  Within session~1 the
pass-to-pass spread (sample standard deviation $0.15$) indicates
unmodeled between-pass variation that may include source, calibration,
compilation, phase, or readout effects; the acquisition does not
identify the source, and the pass-to-pass changes were not monotonic
in the platform-reported HOM value ($92.0\% \to 90.3\%$).  Session~2,
acquired after a same-day recalibration at a steady reported HOM of
$91.5\%$, showed a tighter pass-to-pass spread (sample standard
deviation $0.03$).  Beside
each raw per-pass value we report the fixed-ratio efficiency-reweighted
per-pass estimate $S^{\rm rw}$ of Table~\ref{tab:chsh}, as an
archived-metadata model scenario for the dominant identified
systematic: once it is included, the weakest reweighted pass
(session~1, pass~0, $2.040 \pm 0.062$) overlaps $S=2$ within $1\sigma$.
We therefore report the reweighted series alongside the raw pass-level
values as a sensitivity to the dominant identified readout imbalance.
The between-pass dispersion, not any single
count-pooled shot-noise error, is the pass-level scale; exploratory
pass-level Hartung--Knapp sensitivity errors
(HK SE $0.048$ raw, $0.060$ reweighted) are reported in
Appendix~\ref{app:stats}.

\textit{Count-pooled descriptors.}  Pooling all
$26{,}833$ accepted coincidences gives
\begin{equation}
\label{eq:result}
  S_{\rm count} = 2.485 \pm 0.019,
\end{equation}
where the uncertainty describes conditional counting noise.  Because
the passes show excess dispersion, this is not a pass-level precision
estimate.  The count-pooled session values were
$S = 2.397 \pm 0.028$ ($13{,}551$ coincidences) and
$S = 2.575 \pm 0.027$ ($13{,}282$ coincidences).  The combined
correlators [Eq.~\eqref{eq:E}] were $E(a_0,b_0) = +0.633$,
$E(a_0,b_1) = +0.629$, $E(a_1,b_0) = +0.639$, and
$E(a_1,b_1) = -0.584$ on $5{,}899$--$7{,}162$ coincidences each; the
per-setting counts are tabulated in Appendix~\ref{app:counts}.  The
fixed-ratio reweighted count-pooled value $S^{\rm rw}_{\rm count}$ in
Table~\ref{tab:chsh} is analyzed below.

\textit{Heterogeneity.}  The passes show excess dispersion relative to
their conditional counting errors.  Session~1 scattered about its mean
with $Q/\nu = 21.7/3 = 7.2$, a clear excess dispersion indicating
unmodeled between-pass variation.
Session~2 showed no detectable excess dispersion relative to the
conditional counting errors ($Q/\nu = 1.1/3 = 0.36$).  For the
eight passes combined, $Q/\nu = 42.4/7 = 6.1$.  We therefore
treat the pass as the analysis unit and report per-pass values with
the session means and sample standard deviations above.
Exploratory random-effects and Birge--Student constructions on the
eight passes are collected in Appendix~\ref{app:stats} as sensitivity
analyses under an exchangeability working model that the design does
not establish.

\textit{Operational comparison with $S=2$.}  All
eight raw pass estimates exceeded $S=2$, with the weakest $3.5$
conditional shot-noise standard deviations above that value.  Because
the common-state, local-analyzer, and setting-independent-postselection
premises are not established (Table~\ref{tab:nosig}), this is an
operational comparison rather than an entanglement-witness
significance.  After reweighting, the weakest pass is
$2.040\pm0.062$ and is statistically consistent with $S=2$ at this
conditional precision.

\textit{Readout-efficiency reweighting (fixed-ratio sensitivity).}
The dominant identified non-sign-definite systematic is the
transmittance imbalance of the block-$A$ logical output ports: the
platform-reported relative transmissions at session start were
$119.99$ versus $78.75$ in session~1
($\kappa_{A,1}=1.524$) and $104.98$ versus $66.15$ in session~2
($\kappa_{A,2}=1.587$); the corresponding block-$B$ ratios are
$0.994$ and $0.968$ (Appendix~\ref{app:readout}).  The platform
interface does not return contemporaneous post-compilation
port-calibration uncertainties, so we treat each session's
$\kappa_{A}$ as an assumed constant applied to logical outcome-$1$
counts after server-side compilation rather than as a calibrated
correction, and we report fixed-ratio reweighted sensitivity
descriptors with the same separation of units used for the raw
analysis.  As a conditional count-pooled descriptor, reweighting the
block-$A$ outcome-$1$ counts with these session-specific ratios
lowered the combined value to
\begin{equation}
\label{eq:result_corr}
  S^{\rm rw}_{\rm count} = 2.380 \pm 0.021 \ \text{(shot noise)},
\end{equation}
with the reweighted shot-noise error obtained by the multinomial
delta method (Appendix~\ref{app:readout}).  Taking a
$\pm10\%$ relative excursion about the archived session ratios gives
\begin{equation}
\label{eq:kappa_band}
  S^{\rm rw}_{\rm count} \in [\,2.356,\,2.406\,]
  \ \text{(narrow $\kappa$ band, $\pm10\%$)},
\end{equation}
a systematic comparable to the conditional shot noise and one we do
not add in quadrature to it; separately, the marginal-consistent
value $\kappa_\eta = 1.395$ that zeroes the $b_1$ remote-setting
difference (Appendix~\ref{app:readout}) yields
$S^{\rm rw}_{\rm count} = 2.404$.  The remote-setting diagnostics
below show that a single fixed ratio is mis-specified rather than
merely imprecise: reweighting by the archived $\kappa_A$
flips the sign of the $b_1$ block-$B$ marginal difference (from
$+0.100 \pm 0.008$ to $-0.033 \pm 0.009$) while leaving the $a_1$
block-$A$ difference essentially unchanged
($+0.037 \pm 0.008 \to +0.038 \pm 0.009$).  The $a_1$ residual is
essentially unchanged by the archived-ratio reweight, indicating that
a single fixed ratio does not capture the full setting dependence.
To bracket this mis-specification we report, as an
\emph{ad hoc} stress scan and not a calibrated uncertainty band, the
range spanned by equal-session ratios $\kappa\in[1.2,1.8]$
applied to both sessions,
\begin{equation}
\label{eq:kappa_band_wide}
  S^{\rm rw}_{\rm count} \in [\,2.341,\,2.441\,]
  \qquad(\kappa\in[1.2,1.8]).
\end{equation}
Applying the session-specific ratios to both blocks
($\kappa_{B,1}=0.994$, $\kappa_{B,2}=0.968$) gives
$S^{\rm rw}_{AB} = 2.376 \pm 0.021$, only $0.004$ below the
block-$A$-only value.  Per-pass reweighted values under
session-specific ratios are listed in Table~\ref{tab:chsh}
(range $2.040$--$2.531$); the weakest reweighted pass
(session~1, pass~0, $2.040 \pm 0.062$) overlaps $2$ within $1\sigma$.
The platform reports no
uncertainty on its transmission metadata, so we propagate the ratios
as a sensitivity analysis rather than as calibrated error bars; each
ratio was recorded at session start and was not re-measured during or
after that session, so contemporaneous port-efficiency calibration is
a required upgrade for future runs (Sec.~\ref{sec:limits}).

\begin{table}[tbp]
\centering\scriptsize
\caption{Remote-setting marginal differences
(no-signaling-style diagnostic)}
\label{tab:nosig}
\setlength{\tabcolsep}{2.5pt}
\begin{tabular}{@{}lcc@{}}
\toprule
Comparison & Raw ($z$) & Reweighted ($z$) \\
\midrule
$P(A{=}0)$, $a_0$ & ${-}0.004\,{\pm}\,0.009$ ($-0.5$)
  & ${-}0.005\,{\pm}\,0.009$ ($-0.5$) \\
$P(A{=}0)$, $a_1$ & ${+}0.037\,{\pm}\,0.008$ ($+4.5$)
  & ${+}0.038\,{\pm}\,0.009$ ($+4.4$) \\
$P(B{=}0)$, $b_0$ & ${+}0.010\,{\pm}\,0.008$ ($+1.2$)
  & ${+}0.010\,{\pm}\,0.009$ ($+1.1$) \\
$P(B{=}0)$, $b_1$ & ${+}0.100\,{\pm}\,0.008$ ($+12.0$)
  & ${-}0.033\,{\pm}\,0.009$ ($-3.9$) \\
\bottomrule
\end{tabular}
\begin{minipage}{\linewidth}\vspace{4pt}\footnotesize\noindent\emph{Notes.}~
Differences in one block's outcome marginal between the two settings
of the other block, signed as
$P(\cdot{=}0\,|\,\mathrm{setting}_1)-P(\cdot{=}0\,|\,\mathrm{setting}_2)$
with $(\mathrm{setting}_1,\mathrm{setting}_2)=(a_0b_0,a_0b_1)$ at fixed
$a_0$, $(a_1b_0,a_1b_1)$ at fixed $a_1$, $(a_0b_0,a_1b_0)$ at fixed
$b_0$, and $(a_0b_1,a_1b_1)$ at fixed $b_1$, from the pooled counts of
Table~\ref{tab:marginals}; binomial uncertainties; $z$ is the difference
in units of its standard error.  The reweighted columns apply the
session-specific fixed-ratio reweights $\kappa_{A,1}=1.524$ and
$\kappa_{A,2}=1.587$ to the block-$A$ outcome-$1$ counts
(Appendix~\ref{app:readout}).  Under a
common-ensemble, local-analyzer model with setting-independent
postselection these differences would vanish up to statistics.
\end{minipage}
\end{table}

\textit{Marginal diagnostics.}  Under the common-ensemble,
local-analyzer model, one block's outcome marginal must not depend on
the other block's setting.  Table~\ref{tab:nosig} reports the
four remote-setting comparisons with uncertainties.  Two are
statistically consistent with zero at this precision, whereas two show
clear deviations.  The block-$B$ marginal at fixed $b_1$
shifted by $+0.100 \pm 0.008$ between the $a_0$ and $a_1$ settings
($z \approx 12$), and the block-$A$ marginal at fixed $a_1$ shifted
by $+0.037 \pm 0.008$ ($z \approx 4.5$); both signs were consistent
across all eight passes (per-pass means $+0.100$ and $+0.037$,
sample standard deviations $0.038$ and $0.021$).  The larger
discrepancy is largely accounted for by the readout imbalance:
because acceptance conditions on one click per block, the asymmetric
block-$A$ efficiency distorts the accepted block-$B$ marginals, and
reweighting by the session-specific $\kappa_{A}$ ratios moved the
$b_1$ difference to
$-0.033 \pm 0.009$ ($z \approx -3.9$), passing through zero to an
opposite-sign residual and thereby showing that a single fixed
efficiency ratio is mis-specified, not merely incomplete.  The block-$A$ difference at
$a_1$, in contrast, was essentially unchanged by the reweight
($+0.038 \pm 0.009$ reweighted) and remains unexplained; candidate
contributors include outcome-dependent postselection beyond the
modeled efficiency channel, setting-dependent compilation,
within-pass drift combined with the fixed setting order (below), or
cross-block mixing in the executed unitary.  These candidates are not
separable in the present fixed-order, single-mapping data set; two
control acquisitions that would discriminate among them are listed
as required upgrades in Sec.~\ref{sec:limits}: the same
target matrix submitted under swapped setting labels, and one setting
repeated at two different positions within a pass.  The residual
setting dependencies therefore prevent the present data from
validating the common-ensemble, local-analyzer,
setting-independent-postselection model.  The $z$-scores of
Table~\ref{tab:nosig} are conditional shot-noise diagnostics under
the fixed accepted ensemble.

\textit{Acquisition-order analysis.}  Within every pass the four
settings were acquired in the fixed order $(a_0,b_0)$, $(a_0,b_1)$,
$(a_1,b_0)$, $(a_1,b_1)$, so setting identity and acquisition
position are perfectly confounded.  The per-pass, per-position
correlators and accepted fractions are tabulated in
Table~\ref{tab:perpos}, whose columns are labeled by both position and
the setting that always occupied it.  A linear
additive drift does not, in general, cancel in the CHSH combination:
for a drift component $d_i = \beta i$ at positions $i = 1,\dots,4$,
its contribution to the assembled score is
$d_1 + d_2 + d_3 - d_4 = 2\beta$, which is nonzero.  Modeling the
observed per-position correlator in pass/session $s$ as
\begin{equation}
\label{eq:driftmodel}
  \widetilde{E}_{s,i} = E^{(0)}_{s,i} + \beta_s\,i,
  \qquad i = 1,\dots,4,
\end{equation}
where $E^{(0)}_{s,i}$ is the unknown drift-free correlator of the
setting always acquired at position $i$, yields
\begin{equation}
\label{eq:driftshift}
  S_{\rm obs,s} = S_{\rm true,s} + 2\beta_s.
\end{equation}
Because each setting always occupies one acquisition position,
$\beta_s$ is not identifiable without assumptions on the four
drift-free correlators.  No model-independent drift correction can
therefore be extracted from this dataset.  Randomized or
counterbalanced setting orders (for example, balanced Latin squares
across passes) would break this setting-position degeneracy in future
runs (Table~\ref{tab:portspec}).

\textit{Photon-recycling comparison.}  A two-job A/B pair (setting
$(a_0,b_0)$) found no resolvable effect of the platform's
photon-recycling mitigation on the correlator at this precision
($\Delta E = 0.055 \pm 0.041$); this is an underpowered two-job test,
so it neither establishes nor excludes a small effect.  Recycling was
disabled for the primary data primarily on auditability grounds,
because event reconstruction alters postselected two-photon statistics
in a way that is difficult to audit, not because of measured harm, and the
full comparison is archived in the reproducibility package~\cite{phaseb_zenodo}.

\section{Systematics and limitations}
\label{sec:limits}

\textit{Accepted logical fraction.}  The accepted logical fraction
among returned two-photon samples is $0.147$--$0.179$ per setting
(Appendix~\ref{app:counts}).  Because its denominator is already
conditioned on the provider's detected-photon filter, it is a
descriptive conditional acceptance rate rather than an empirical
estimate of the ideal $1/9$ gate-success probability.
The setting dependence of this rate nonetheless bears
on the setting-independent-postselection assumption, so we compare it
with the archived targets.  Evaluating the two-photon output
distributions of the four submitted $16\times16$ unitaries and
conditioning on the same two-distinct-click denominator, the ideal
coincidence-basis acceptance is exactly $1/9$ for every setting, and
the accepted fraction among two-click patterns is predicted to be
$0.154$ at $(a_0,b_0)$ and $(a_1,b_0)$ and $0.133$ at $(a_0,b_1)$ and
$(a_1,b_1)$ (at $v=1$; $0.161$ and $0.141$ at the fitted $v=0.94$,
$0.258$ and $0.242$ in the fully distinguishable limit).  The measured
fractions ($0.147$, $0.175$, $0.169$, $0.179$ in acquisition
order) do not follow that pattern: the model predicts a dependence on
the block-$B$ setting alone, with $(a_1,b_1)$ the lowest of the
four, whereas the data have it highest and increase almost
monotonically with acquisition position.  The measured setting
dependence of acceptance is therefore not explained by the submitted
targets under either the coherent or the fitted-contrast model, and is
consistent with the acquisition-order confounding discussed below; the
comparison is regenerated by \texttt{verify\_targets.py} in the
reproducibility package~\cite{phaseb_zenodo}.  The observed pattern
is not captured by the submitted-target model.
Separately, contrast loss from imperfectly
interfering events dilutes the correlators toward zero and is the
dominant gap between the measured $S$ (Eq.~\eqref{eq:result}) and the
ideal $2\sqrt2 \approx 2.83$.  We model that loss next without
appeal to the accepted-fraction value.

We describe the contrast loss with a two-component mixture of the
exact coherent two-photon output distribution and its fully
distinguishable counterpart, both computed from the archived target
unitaries generated by the circuit-construction code,
\begin{equation}
\label{eq:vmodel}
  P(v) = v\,P_{\mathrm{coh}} + (1-v)\,P_{\mathrm{dist}},
\end{equation}
where $P_{\mathrm{coh}}$ is the two-photon detection distribution for
fully indistinguishable photons, $P_{\mathrm{dist}}$ that for
orthogonal (fully distinguishable) photons, and $v \in [0,1]$ is a
descriptive effective-contrast parameter fitted from the CHSH data
themselves; it should not be interpreted as an independent
measurement of photon indistinguishability.  The model predicts an
operational $S>2$ for $v \gtrsim 0.83$ and
$S = 2\sqrt2$ at $v = 1$.  Matching the scalar CHSH value to this
one-parameter curve gives $v \approx 0.92$ (session~1), $0.95$
(session~2), and $0.94$ (combined).  The reproducibility package also
includes a least-squares fit to the four archived session~1
correlators, which returns $v \approx 0.93$ and illustrates the
per-setting asymmetry not captured by a scalar $S$ match.
Within session~1,
raw $S$ rose from $2.202$ to $2.541$ while the dashboard HOM fell
from $92.0\%$ to $90.3\%$.  The opposing trends show that $v$ cannot
be identified with the dashboard HOM visibility alone; separating
source indistinguishability from interferometric phase drift would
require independent phase monitoring.  The fitted $v \approx 0.91$--$0.95$
is of similar scale to the platform-reported HOM visibility of
$91$--$92\%$, paralleling the
single-block observation of Ref.~\cite{q3paper} that visibility
degradation removed the neutral-sector suppression, which
recalibration then restored.

\begin{table*}[tbp]
\caption{Uncertainty and contrast-loss summary}
\label{tab:syst}
\centering\footnotesize
\begin{tabularx}{\textwidth}{@{}l>{\raggedright\arraybackslash}X>{\raggedright\arraybackslash}X@{}}
\toprule
Source & Effect on $S$ & Character \\
\midrule
\multicolumn{3}{@{}l}{\emph{Inferential uncertainty}}\\
Raw count-pooled       & $2.485\pm0.019$  & shot noise \\
Raw pass-level         & HK SE $0.048$  & sensitivity \\
Reweighted count-pooled & $2.380\pm0.021$ & model scenario \\
Reweighted pass-level  & HK SE $0.060$  & sensitivity \\
Readout imbalance      & $-0.10$ (reweight.)  & App.~\ref{app:readout} \\
$\kappa_\eta$ narrow band & $2.356$--$2.406$ & Eq.~\eqref{eq:kappa_band} \\
$\kappa_\eta$ \emph{ad hoc} scan & $2.341$--$2.441$ & Eq.~\eqref{eq:kappa_band_wide} \\
$\kappa_\eta$ uncert.  & not reported     & platform \\
\midrule
\multicolumn{3}{@{}l}{\emph{Modeled contrast loss} (descriptive)}\\
Residual gap to $2\sqrt2$ & $-0.34$ ($v\approx0.94$) & fitted match \\
\quad phase effects   & folded into $v$ & sign not separately identified \\
Multiphoton ($g^{(2)}$) & unquantified & not a signed bound \\
Dark-count accidentals & two-dark negligible & mixed events unquantified \\
Common-mode loss      & $0$             & cancels in $E$ \\
Leakage outside $\mathcal{Q}_L$ & removed & lower accept. \\
\midrule
\multicolumn{3}{@{}l}{\emph{Structural/model assumptions} (no error bar)}\\
Common prep $+$ analyzers & assumed & mapping n/a \\
Setting-indep.\ postsel.\   & assumed & Tab.~\ref{tab:nosig} \\
Residual marginals ($a_1$) & model tension & $+0.038$ \\
Fair sampling (imbal.\ $\eta$) & assumed & sensitivity \\
Fixed order; serial passes      & confounds & no drift bound \\
$\kappa_\eta$ stab.; leakage & unmeasured & structural \\
\bottomrule
\end{tabularx}
\begin{minipage}{\linewidth}\vspace{4pt}\footnotesize\noindent\emph{Notes.}~
The three groups are inferential uncertainties (upper), modeled
contrast-loss estimates (middle), and structural assumptions (lower);
they are not combined into a single additive error budget.
Count-pooled descriptors are kept separate from the exploratory
pass-level Hartung--Knapp (HK) sensitivity errors in
Appendix~\ref{app:stats}.  The $\kappa$ intervals are sensitivity
ranges rather than statistical intervals and are not added in
quadrature to the conditional errors.  The multiphoton channel is
unquantified, and mixed one-real plus one-dark events are not
separately bounded.
\end{minipage}
\end{table*}

\textit{Error budget.}  Table~\ref{tab:syst} separates the
inferential uncertainties on $S$ from the modeled contrast-loss
mechanisms, which we do not combine into a single additive budget.
Two structural features nonetheless limit the effect of
several of these channels.
First, the correlator of Eq.~\eqref{eq:E} is a normalized ratio, so
any loss that is outcome independent, setting independent, and
multiplicative within a correlator (the common-mode transmittance,
which dominates the $\sim\!5\%$ per-photon budget) cancels in $E$.
Second, in the simplified models used here, partial
distinguishability [Eq.~\eqref{eq:vmodel}] and the modeled
background channels (residual multiphoton emission and detector
dark counts) reduce the correlation contrast.  We
do not assert sign-definiteness in general: setting-dependent filtering,
multiphoton events through the global interferometer, and dark
counts combined with asymmetric postselection need not be
sign-definite without a model, and outcome-dependent readout loss
(below) is treated separately.  Quantitatively, at $\le 52~\mathrm{s}^{-1}$ and a $4.94$~MHz clock
the dark-count probability is at most $1.1\times10^{-5}$ per detector
per clock cycle.  The probability of a purely two-dark coincidence
across the four cross-block logical-port pairs is therefore of order
$4\times10^{-10}$ per clock cycle.  This is negligible relative to the
nominal $(0.05)^2/9 \approx 2.8\times10^{-4}$ accepted-coincidence
scale.  Mixed events containing one genuine detection and one dark
count depend on the detailed detection and postselection model and are
not separately bounded here.  Residual multiphoton emission is estimated
at the order-of-magnitude level by a lost-twin scaling under a
low-multiphoton source model: the quantum-dot source feeds a
temporal-to-spatial demultiplexer that routes successive pulses to the
two block inputs, so a spurious extra photon requires a two-photon
Fock component in one pulse (relative two-photon weight of order
$\tfrac12 g^{(2)}(0) \approx 5\times10^{-3}$) with a twin
surviving the $\sim\!5\%$ transmission and firing a threshold
detector.  The lost-twin estimate is therefore order-of-magnitude
only; without a Fock/threshold simulation through the target
unitaries, we do not assign this channel a signed bias on $S$.
The non-sign-definite channel for
which we perform an explicit quantitative sensitivity analysis is
outcome-dependent readout loss: the archived output
transmittances of the two logical ports of block~$A$ differ by
session-specific factors $\kappa_{A,1}=1.524$ and
$\kappa_{A,2}=1.587$ (block-$B$ ratios $0.994$ and $0.968$), an
imbalance the fair-sampling assumption leaves uncorrected.  Because $E$ is a
normalized correlator this imbalance would cancel for balanced
block-$A$ outcome marginals, but the marginals are in fact strongly
imbalanced, so reweighting the block-$A$ outcomes by the
session-specific start-of-session ratios shifts the combined value
down by $\approx0.10$ to
$S^{\rm rw} = 2.380 \pm 0.021$ (Sec.~\ref{sec:results};
Appendix~\ref{app:readout}).  Whether the true
outcome-dependent filtering is captured by session-fixed ratios is
not established by these data.  An exploratory pass-level sensitivity
analysis is given in Appendix~\ref{app:readout}.  Its full removal would require efficiency-calibrated
readout, and even then would not close the detection loophole; both
are deferred to future work.

\textit{Assumptions and scope.}  The operational quantity $S$ is well
defined from the accepted four-way counts
[Eqs.~\eqref{eq:E}--\eqref{eq:S}].  Interpreting $S>2$ as an
entanglement witness additionally requires a common bipartite state,
block-local analyzers, and sufficiently setting-independent
acceptance; the marginal diagnostics of Table~\ref{tab:nosig} do not
establish these premises.  At the present effective detection
efficiency, interpreting the postselected correlations as
representative of the underlying ensemble additionally relies on a
fair-sampling assumption~\cite{pearle1970,clauser1974,garg1987,eberhard1993}.

Both qubits are measured on one processor with software-defined
settings and are not space-like separated.  The present measurement is
therefore a postselected reference acquisition rather than a
device-independent or loophole-free Bell
test~\cite{hensen2015,giustina2015,shalm2015}.

\textit{Limitations of compilation auditability.}  The offline checks
certify the submitted target matrices
(Table~\ref{tab:hashes}; Eq.~\eqref{eq:t4residual}) but not the
server-side compiled transformation, which is not returned.  The
measured acceptance pattern and residual remote-setting marginals do
not independently verify preservation of the factorization in
Eq.~\eqref{eq:factorization}.  A setting-dependent compiled
transformation could therefore alter the interpretation of $S$ while
leaving the operational score on returned counts well defined.

\begin{table}[tbp]
\caption{Status of the principal claims}
\label{tab:claims}
\centering\footnotesize
\setlength{\tabcolsep}{3pt}
\begin{tabular}{@{}p{0.40\columnwidth}p{0.24\columnwidth}p{0.26\columnwidth}@{}}
\toprule
Claim & Status & Basis \\
\midrule
Submitted target unitaries & Verified &
  hashes; App.~\ref{app:cz} \\
Operational $S$ on accepted counts & Verified &
  Eqs.~\eqref{eq:E}--\eqref{eq:S} \\
Executed compiled mapping & Not established &
  never returned \\
Common-state / local-analyzer model & Not established &
  Table~\ref{tab:nosig} \\
Setting-independent postselection & Not established &
  Table~\ref{tab:nosig} \\
Fair sampling at ${\sim}5\%$ efficiency & Assumed &
  Sec.~\ref{sec:limits} \\
Entanglement-witness status & Not established &
  Table~\ref{tab:nosig} \\
Error-detected logical Bell pair & Not demonstrated &
  no syndrome \\
$\kappa$ temporal stability & Assumed &
  App.~\ref{app:readout} \\
Provider \texttt{min\_detected\_photons} filter & Not characterized &
  Sec.~\ref{sec:limits} \\
\bottomrule
\end{tabular}
\begin{minipage}{\linewidth}\vspace{4pt}\footnotesize\noindent\emph{Notes.}~
Interpretive status of each principal claim.  ``Verified'' means
established by the archived offline checks or by the returned counts;
``Assumed'' means adopted as a working premise not independently
established here; ``Not established''/``Not characterized''/``Not
demonstrated'' mean the present data do not support the statement.
\end{minipage}
\end{table}

\textit{Provider photon-filter selection stage.}  A distinct
fair-sampling concern enters upstream of our offline postselection
through the provider's own returned-sample filter.  Every returned
pattern has already passed the Perceval \texttt{min\_detected\_photons}
filter (Sec.~\ref{sec:protocol}), applied by the remote processor
before any pattern is delivered; the detailed detector/filter
implementation and any dead-time or event-reconstruction behavior are
not exposed through the
interface and were not characterized by us.  This filter is therefore
an uncharacterized prior selection stage that sits between the physical
detection events and the returned samples on which every rate and
correlator we quote is conditioned, and its interaction with the
block-$A$ efficiency imbalance and with any setting-dependent detection
cannot be audited from the returned data.

\textit{Reference-acquisition specification and hardware requirements.}
Table~\ref{tab:portspec} summarizes the portable reference-acquisition
specification and identifies the elements absent from the present
runs.

\begin{table*}[tbp]
\caption{Portable CHSH reference-acquisition specification}
\label{tab:portspec}
\centering\footnotesize
\begin{tabular*}{\textwidth}{@{\extracolsep{\fill}}c p{0.72\textwidth} c@{}}
\toprule
Item & Specification & Satisfied here? \\
\midrule
(i)   & Four target $16\times16$ unitaries generated by
        \texttt{logical\_2block\_chsh\_belenos.py}, identified by the
        SHA-256 content hashes of Table~\ref{tab:hashes} & Yes \\
(ii)  & Logical port list (block $A$ modes $0$--$7$, block $B$ modes
        $8$--$15$; logical readouts after rail extraction) & Yes \\
(iii) & Perceval/\texttt{Sampler} parameters, including
        \texttt{max\_samples}${}=5000$ and the physical
        detected-photon filter & Yes \\
(iv)  & Offline postselection predicate (exactly one logical-port
        click per block) & Yes \\
(v)   & Estimators $S_{\rm count}$, $S_{\rm pass}$, and fixed-ratio
        $S^{\rm rw}_{\rm count}$ & Yes \\
(vi)  & Efficiency reporting: report raw plus fixed-ratio reweight with a
        declared $\kappa_{A}$ source & Yes \\
(vii) & Contemporaneous post-compilation port calibration with reported
        uncertainties (preferred upgrade when available) & No \\
(viii)& Randomized or counterbalanced setting order across passes & No \\
(ix)  & Rate-denominator convention (returned samples after the
        physical photon filter, not raw source-clock shots) & Yes \\
(x)   & Required outputs (four-way counts, job IDs, timestamps,
        per-pass $S$, marginal diagnostics) & Yes \\
(xi)  & Reporting-standard status: no pass/fail threshold is defined;
        comparison uses the archived reference under this protocol & Yes \\
(xii) & Contemporaneous characterization (or disabling) of the provider
        \texttt{min\_detected\_photons} returned-sample filter & No \\
\bottomrule
\end{tabular*}
\begin{minipage}{\linewidth}\vspace{4pt}\footnotesize\noindent\emph{Notes.}~
Self-contained specification for transporting the protocol to another
provider.  ``Satisfied here?'' records whether the present
fixed-order reference runs meet each item: items (vii), (viii), and (xii)
are not satisfied: the runs used start-of-session $\kappa_{A}$
metadata rather than contemporaneous port calibration, a fixed
(non-counterbalanced) setting order, and an uncharacterized provider
returned-sample filter.  Items (i)--(vi) and (ix)--(xi)
are satisfied.  Minimum hardware prerequisites are two input photons,
at least $16$ addressable modes for this encoding (fewer for a
dual-rail variant), user-specified sufficiently general linear-optical
unitaries, raw output-pattern counts, error mitigation disabled or
fully documented, and timestamped jobs with stable identifiers.
\end{minipage}
\end{table*}

\noindent The diagnostics identify the required
upgrades for any replication seeking stronger claims: counterbalanced
or randomized setting order across passes; contemporaneous
post-compilation output-port calibration with reported uncertainties,
replacing start-of-session $\kappa_A$ metadata; contemporaneous
characterization (or disabling) of the provider
\texttt{min\_detected\_photons} returned-sample filter; and, to
identify the residual block-$A$ marginal at $a_1$
(Table~\ref{tab:nosig}), control acquisitions submitting the same
target matrix under swapped setting labels and repeating one setting
at two positions within a pass.  The present dataset can serve as a
like-for-like reference for repetitions using the same target and
acquisition order, but cross-platform comparison requires the upgraded
specification of Table~\ref{tab:portspec}.

\section{Conclusion}
\label{sec:conclusion}

Two spatial-mode qubits, each a two-dimensional neutral-rail subspace
embedded in the zero-sum sector of an eight-mode single-photon
register, were coupled by a target postselected linear-optical
controlled-$Z$ circuit on a commercial cloud photonic processor,
operated end-to-end by external users through the public interface.
Across eight complete CHSH passes in two same-day sessions, the
pass-level means were $2.40$ and $2.58$ with sample standard
deviations $0.15$ and $0.03$, and the passes showed strong excess
dispersion relative to conditional shot noise ($Q/\nu=6.1$).  The raw
count-pooled descriptor was $S_{\rm count}=2.485\pm0.019$; the
fixed-ratio efficiency-reweighted sensitivity scenario gave
$S^{\rm rw}_{\rm count}=2.380\pm0.021$, the weakest reweighted
pass was $2.040\pm0.062$, and the $\kappa$-stress scan spanned
$2.341$--$2.441$.

This result establishes an auditable, fixed-order postselected CHSH
reference acquisition on a commercial cloud photonic processor.
Because the executed compiled mapping is not returned, setting order
was fixed, and residual remote-setting marginal differences remain, the
present data do not establish a common-state, local-analyzer
entanglement-witness interpretation and should not be read as a
loophole-free Bell test or a validated cross-platform performance
metric.  The complete count records, submitted-target hashes, and
analysis pipeline are openly archived to support independent
verification and future like-for-like
replication~\cite{phaseb_zenodo}.  Future work should combine
counterbalanced acquisition order, contemporaneous efficiency
calibration, and independent-day or second-provider replication, while
extending the approach toward lower-loss and multi-step logical
photonic circuits.

\begin{acknowledgments}
The cloud demonstration was performed on Quandela's cloud-accessible Belenos
processor through the Perceval software framework, accessed via the
publicly available Explorer offer on the same terms as any external
user; no dedicated hardware access or dedicated technical support
was provided.  We thank Bhaskar Roy Bardhan for helpful comments on
the manuscript.  This research received no external funding.
\end{acknowledgments}

\noindent\textit{Author contributions.}
E.T.: conceptualization, data curation, formal analysis (lead),
investigation, methodology, software, and writing (original draft).
J.W.: investigation, methodology, software, and supervision.
M.S.: formal analysis, project administration, validation,
visualization, and writing (review and editing).  All authors read
and approved the final manuscript.

\noindent\textit{Conflict of interest.}
The authors declare no conflicts of interest.

\noindent\textit{Data availability.}
The complete count records, cloud-job manifest, calibration metadata
used in the analysis, circuit-construction code, and analysis software
supporting the findings of this article are openly available in the
Zenodo repository cited in Ref.~\cite{phaseb_zenodo}
(DOI
\href{https://doi.org/10.5281/zenodo.21878219}{\mbox{\nolinkurl{doi:10.5281/zenodo.21878219}}}).
The archived package includes a pinned software environment and
verification scripts that reproduce the reported tables and numerical
results without live cloud access.  The verification entry point
\texttt{cd scripts \&\& python3 verify\_targets.py} recomputes the
full $64$-hex SHA-256 digests of the four submitted target unitaries
and of the offline postselection predicate (complete \texttt{classify}
source together with \texttt{PORT\_A}/\texttt{PORT\_B}).  The complete
digests, not only the $16$-hex prefixes of Table~\ref{tab:hashes}, are
archived in the package (\texttt{SHA256SUMS.txt}).  The same check
confirms that the tabulated matrices match the archived construction
under the documented $12$-decimal canonical serialization of
\texttt{logical\_2block\_chsh\_belenos.py}, regenerates
$S_{\rm count}$ and $S^{\rm rw}_{\rm count}$ from the archived counts,
and exits with a nonzero status on any
mismatch.

\appendix

\section{Reduction to the single-mode controlled-\texorpdfstring{$Z$}{Z}}
\label{app:cz}

We show that coupling the two-mode rails sub-mode-wise with identical
beam splitters reproduces the single-mode
Ralph--Langford--Bell--White (RLBW) controlled-$Z$~\cite{ralph2002,%
obrien2003} exactly, so that the postselected action on the logical
subspace is $\tfrac13\,\mathrm{diag}(1,1,1,-1)=\tfrac13\,
\mathrm{CZ}_L$ with success probability $1/9$.

Introduce, for the block-$A$ target rail (modes $2,3$) and block-$B$
target rail (modes $10,11$), the neutral (rail) creation operators
\begin{equation}
\label{eq:railops}
  d_A^\dagger = \frac{a_2^\dagger - a_3^\dagger}{\sqrt2}, \qquad
  d_B^\dagger = \frac{a_{10}^\dagger - a_{11}^\dagger}{\sqrt2},
\end{equation}
which carry the logical amplitude of $|1\rangle_L$ in each block
[Eq.~\eqref{eq:logical}].  The gate couples the sub-modes pairwise by
beam splitters of transmissivity $T$, $a_2\!\leftrightarrow\!a_{10}$
and $a_3\!\leftrightarrow\!a_{11}$, each mapping an input operator to
$\sqrt{T}\,(\cdot) + i\sqrt{\mathcal{R}}\,(\cdot')$ with $\mathcal{R}=1-T$ and
$(\cdot')$ the partner sub-mode.  Because the two beam splitters are
identical and act on disjoint pairs, they preserve the antisymmetric
combination:
\begin{equation}
\label{eq:railBS}
  d_A^\dagger \mapsto \sqrt{T}\,d_A^\dagger + i\sqrt{\mathcal{R}}\,d_B^\dagger,
  \quad
  d_B^\dagger \mapsto i\sqrt{\mathcal{R}}\,d_A^\dagger + \sqrt{T}\,d_B^\dagger,
\end{equation}
while the symmetric combinations
$(a_2^\dagger+a_3^\dagger)/\sqrt2$ and
$(a_{10}^\dagger+a_{11}^\dagger)/\sqrt2$ transform identically among
themselves, decoupled from Eq.~\eqref{eq:railBS}.  Hence in the rail
(difference) sector the two-mode-rail coupling is exactly a
single beam splitter of transmissivity $T$ between the collective modes
$d_A,d_B$: the single-mode RLBW configuration.

For the $|11\rangle_L$ input $d_A^\dagger d_B^\dagger|0\rangle$,
Eq.~\eqref{eq:railBS} yields
\begin{equation}
  d_A^\dagger d_B^\dagger \mapsto
  i\sqrt{T\mathcal{R}}\,\bigl(d_A^{\dagger 2} + d_B^{\dagger 2}\bigr)
  + (T-\mathcal{R})\,d_A^\dagger d_B^\dagger ,
\end{equation}
whose coincidence term (one photon in each block's target rail)
carries amplitude $T-\mathcal{R} = -\tfrac13$ at $T=\tfrac13$; the bunching
terms place both photons in one block and are removed by the
one-click-per-block postselection.  For the other three logical
inputs exactly one photon occupies a target rail, so no two-photon
interference occurs, and the amplitude-balancing dump couplings on
the rail-$0$ sub-modes ($a_0\!\leftrightarrow\!a_4$,
$a_1\!\leftrightarrow\!a_5$; $a_8\!\leftrightarrow\!a_{12}$,
$a_9\!\leftrightarrow\!a_{13}$, also of transmissivity $T$) reduce by
the same argument to single $T$ beam splitters on the rail-$0$
difference modes, so a photon retained in its rail acquires amplitude
$\sqrt{T}=1/\sqrt3$.  Each of $|00\rangle_L$, $|01\rangle_L$,
$|10\rangle_L$ therefore keeps joint amplitude
$\sqrt{T}\cdot\sqrt{T}=\tfrac13$ with no sign change.

Collecting the four inputs, the postselected logical transfer matrix
is $\tfrac13\,\mathrm{diag}(+1,+1,+1,-1)$: the controlled-$Z$ up to
the common factor $\tfrac13$, whose squared modulus $1/9$ is the
success probability.  This coincides with the single-mode RLBW gate,
establishing the assertion of Sec.~\ref{sec:encoding}.

The reproducibility script
\texttt{logical\_2block\_chsh\_belenos.py} numerically assembles the
four target $16\times16$ matrices and checks unitarity residuals
$<10^{-10}$ (largest observed $7.8\times10^{-16}$) together with the
submitted-target factorization of Eq.~\eqref{eq:factorization}.  The
companion script \texttt{verify\_targets.py} also evaluates the
postselected logical $4\times4$ transfer matrix of the coupling stage
by summing the two-photon coincidence amplitudes on the logical rails,
obtaining
\begin{equation}
\label{eq:t4residual}
  \max\bigl|\,T_L - \tfrac13\,\mathrm{diag}(1,1,1,-1)\,\bigr|
  = 1.7\times10^{-16},
\end{equation}
so the analytic reduction above holds numerically at machine
precision on the submitted matrices.  Table~\ref{tab:hashes} lists the
SHA-256 hashes that identify those four matrices; the same script
recomputes them and exits with a nonzero status on any mismatch.  All
three checks concern the submitted targets only: the
server-side compiled physical mapping is not returned and cannot be
verified this way.

\begin{table}[tbp]
\caption{Content hashes of the four submitted CHSH target unitaries}
\label{tab:hashes}
\centering\footnotesize
\begin{tabular}{@{}ll@{}}
\toprule
Setting & SHA-256 (first 16 hex digits) \\
\midrule
$(a_0,b_0)$ & \texttt{aa2f0f0b98ed81ca} \\
$(a_0,b_1)$ & \texttt{825367db121109f3} \\
$(a_1,b_0)$ & \texttt{c2f86e41d38ff99d} \\
$(a_1,b_1)$ & \texttt{60a9e73c607f8302} \\
\bottomrule
\end{tabular}
\begin{minipage}{\linewidth}\vspace{4pt}\footnotesize\noindent\emph{Notes.}~
SHA-256 digests of the $16\times16$ complex target matrices assembled
by \texttt{logical\_2block\_chsh\_belenos.py}, computed over real and
imaginary parts rounded to $12$ decimals and serialized as
little-endian \texttt{float64} in row-major order (signed zeros
normalized).  Identity under this documented canonical serialization
is what the verification checks.  Only the first $16$ hex digits are
shown for typesetting; the full $64$-hex SHA-256 digests are archived
in the reproducibility package~\cite{phaseb_zenodo}
(\texttt{SHA256SUMS.txt}).  The postselection-predicate digest
(\texttt{6bae83187a6bba37}$\ldots$, Table~\ref{tab:runparams}) hashes
the complete \texttt{classify} source together with
\texttt{PORT\_A}/\texttt{PORT\_B}.
\end{minipage}
\end{table}

\section{Statistical methods}
\label{app:stats}

Unless otherwise stated, uncertainties reported as $\pm$ values for
correlators and CHSH scores denote one conditional shot-noise standard
error.  The Wilson intervals in Table~\ref{tab:tt} are two-sided
$95\%$ confidence intervals.  Pass-level Hartung--Knapp intervals
below are exploratory sensitivity intervals under an exchangeability
working model.  For a correlator estimated from $n$
postselected coincidences via Eq.~\eqref{eq:E}, binomial error
propagation on $E = (N_{\rm corr} - N_{\rm anti})/n$, with
$N_{\rm corr} = N_{00} + N_{11}$ and
$N_{\rm anti} = N_{01} + N_{10}$, gives the conditional shot-noise
variance given the accepted count $n$,
\begin{equation}
\label{eq:sigmaE}
  \sigma_E^2 \simeq \frac{1 - E^2}{n},
\end{equation}
i.e.\ the asymptotic binomial variance of $E$ at fixed $n$; the four
settings being separate acquisitions,
\begin{equation}
\label{eq:sigmaS}
  \sigma_S^2 = \sum_{x,y}\sigma_{E_{xy}}^2 .
\end{equation}
We analyze the pass-level estimates $S_i$, each comprising one
complete four-setting acquisition.  Their
homogeneity, for the $k$ passes pooled into a session or the eight
passes combined, is measured by Cochran's $Q$
statistic~\cite{cochran1954}
\begin{equation}
\label{eq:chi2}
  Q = \sum_i \frac{(S_i - \bar S)^2}{\sigma_i^2},
  \qquad \nu = k-1,
\end{equation}
where $\bar S = \bigl(\sum_i \sigma_i^{-2} S_i\bigr)/
\bigl(\sum_i \sigma_i^{-2}\bigr)$ is the inverse-variance-weighted
pass mean, $\nu$ the degrees of freedom, and
\begin{equation}
\label{eq:sigmaSbar}
  \sigma_{\bar S}
  = \Bigl(\sum_i \sigma_i^{-2}\Bigr)^{-1/2}
\end{equation}
is the standard error of $\bar S$.  Under independent
approximately normal
measurement errors and a common mean, $Q$ is approximately
$\chi^2_\nu$-distributed.  Because the present passes are sequential
and dependent, we use $Q/\nu$ only as a dispersion diagnostic.
The Cochran $Q$ and $Q/\nu$ values quoted here and in
Sec.~\ref{sec:results} (session~1 $7.2$, session~2 $0.36$, combined
$6.1$) use the raw per-pass estimates $S_i$ and their
conditional shot-noise errors $\sigma_i$ of Table~\ref{tab:chsh}.
When $Q > \nu$ the shot-noise error is inflated by the Birge factor
\begin{equation}
\label{eq:birge}
  f_B = \max\!\left(1,\ \sqrt{Q/\nu}\right),
\end{equation}
giving the heterogeneity-aware standard error $f_B\,\sigma_{\bar S}$.
Because $f_B$ is estimated from only $\nu$ degrees of freedom, the
standardized deviation $(\bar S - 2)/(f_B\sigma_{\bar S})$ is referred to a
Student-$t_\nu$ distribution rather than a standard normal: the
$95\%$ interval quoted below uses $t_7$, so a
Gaussian $\sigma$-label formed as $(\bar S - 2)/(f_B\sigma_{\bar S})$
overstates the significance.
A random-effects meta-analysis of the eight pass estimates gives the
same picture in full detail: the DerSimonian--Laird
estimator~\cite{dersimonian1986} gives a between-pass variance
$\tau^2 = 0.0148$, and maximizing the restricted likelihood over
$\tau^2$ (REML) gives $\tau^2 = 0.0155$; with the corresponding
weights $w_i = (\sigma_i^2 + \tau^2)^{-1}$ the pooled estimate is
$\hat\mu = 2.489$, the Hartung--Knapp~\cite{hartung2001} standard
error is
$\bigl[\sum_i w_i (S_i - \hat\mu)^2 / ((k-1)\sum_i w_i)\bigr]^{1/2}
= 0.048$, and the resulting exploratory $t_7$ sensitivity interval is
\begin{equation}
\label{eq:result_scaled}
  S_{\rm pass} = 2.489, \qquad
  [\,2.374,\ 2.603\,]\ \ (t_7),
\end{equation}
under the exchangeability working model.  No meta-analysis package was used: the
DerSimonian--Laird and REML estimators of $\tau^2$, the
Hartung--Knapp standard error given explicitly above, and the
Birge--Student construction are implemented directly in
\texttt{revision\_stats.py} in the reproducibility
package~\cite{phaseb_zenodo}, which is therefore the operative
definition of every quantity in this appendix.  Because the eight passes are
sequential acquisitions on shared hardware, calibration history,
compiler, and source, their independence is not asserted; both the
Birge--Student and the random-effects constructions are reported as
sensitivity analyses describing pass-level dispersion, not as
definitive frequentist tests.  To avoid a hybrid construction, the
count-pooled score $S_{\rm count} = 2.485 \pm 0.019$ is reported as a
conditional descriptor with its own shot-noise error, while the
pass-level interval of Eq.~\eqref{eq:result_scaled} is centered on a
pass-level estimator, namely the random-effects pooled value
$\hat\mu = 2.489$, or equivalently the Birge--Student construction
about the inverse-variance-weighted pass mean $\bar S \approx 2.494$
with error $f_B\,\sigma_{\bar S}$.  The count-pooled, weighted-mean, and
random-effects centers differ by at most $\approx 0.01$, an order of
magnitude below the interval half-width, so the distinction does not
affect any reported figure but keeps the two units of analysis
separate.
Proportions (the conditional correctness probabilities of
Table~\ref{tab:tt}) are reported with $95\%$ Wilson score
intervals~\cite{wilson1927}.

Standard CHSH-to-entanglement conversions are not reported because the
present implementation does not independently establish their
common-state and local-measurement premises
(Sec.~\ref{sec:limits}).

\section{Per-setting CHSH counts}
\label{app:counts}

Table~\ref{tab:persetting} gives, for each of the four combined CHSH
settings, the accepted two-photon coincidences $n$ (pooled over the
eight passes, out of $8\times5{,}000 = 40{,}000$ returned two-photon
samples per setting), the accepted logical fraction $n/40{,}000$ among
those returned samples, the correlated and anticorrelated coincidence
sums $N_{00}+N_{11}$ and $N_{01}+N_{10}$ that form the numerator and
denominator of Eq.~\eqref{eq:E}, the correlator $E$, and its
conditional shot-noise uncertainty $\sigma_E$ [Eq.~\eqref{eq:sigmaE}].
Table~\ref{tab:marginals} gives the corresponding four-way counts and
block marginals used for the readout-imbalance and
no-signaling-style diagnostics.  The four correlators combine to
$S = 2.485 \pm 0.019$
[Eqs.~\eqref{eq:S} and~\eqref{eq:result}].  This fraction uses the
returned-sample denominator (patterns already passing the physical
detected-photon filter) and is not directly comparable to the ideal
coincidence-basis $1/9$ (Sec.~\ref{sec:limits}).  Comparing the
marginals across the remote setting gives
$\Delta P(B{=}0)=+0.010 \pm 0.008$ at $b_0$ and $+0.100 \pm 0.008$ at
$b_1$, and $\Delta P(A{=}0)=-0.004 \pm 0.009$ at $a_0$ and
$+0.037 \pm 0.008$ at $a_1$ (binomial uncertainties); the main text
analyzes these differences, their per-pass dispersion, and their
behavior under the efficiency reweighting
(Table~\ref{tab:nosig}).  These
checks are diagnostics of setting-dependent sampling, not
Bell-loophole closures; the full per-pass four-way counts are
archived in the reproducibility package.

\begin{table*}[tbp]
\caption{Combined CHSH counts by setting}
\label{tab:persetting}
\begin{tabular*}{\textwidth}{@{\extracolsep{\fill}}lcccccc@{}}
\toprule
Setting & $n$ & frac. & $N_{00}{+}N_{11}$ & $N_{01}{+}N_{10}$
        & $E$ & $\sigma_E$ \\
\midrule
$(a_0,b_0)$ & $5899$ & $0.147$ & $4817$ & $1082$ & $+0.633$ & $0.0101$ \\
$(a_0,b_1)$ & $7016$ & $0.175$ & $5715$ & $1301$ & $+0.629$ & $0.0093$ \\
$(a_1,b_0)$ & $6756$ & $0.169$ & $5538$ & $1218$ & $+0.639$ & $0.0094$ \\
$(a_1,b_1)$ & $7162$ & $0.179$ & $1491$ & $5671$ & $-0.584$ & $0.0096$ \\
\midrule
Combined & $26833$ & --- & --- & --- & \multicolumn{2}{c}{$S=2.485\pm0.019$} \\
\bottomrule
\end{tabular*}
\begin{minipage}{\linewidth}\vspace{4pt}\footnotesize\noindent\emph{Notes.}~
Accepted two-photon coincidences $n$ pooled over the eight
passes of both CHSH sessions.  ``frac.'' is the accepted logical
fraction $n/40{,}000$ among returned two-photon samples (denominator
$8\times5{,}000$); it is conditioned on the returned-sample filter and
is not directly comparable to the ideal $1/9$ (Sec.~\ref{sec:limits}).
$N_{00}{+}N_{11}$ and $N_{01}{+}N_{10}$ are the correlated and
anticorrelated coincidences of Eq.~\eqref{eq:E}, and $\sigma_E$ is
the conditional binomial shot-noise uncertainty
[Eq.~\eqref{eq:sigmaE}].
\end{minipage}
\end{table*}

\begin{table*}[tbp]
\caption{Four-way pooled counts and marginal diagnostics}
\label{tab:marginals}
\begin{tabular*}{\textwidth}{@{\extracolsep{\fill}}lrrrrrrrr@{}}
\toprule
Setting & $N_{00}$ & $N_{01}$ & $N_{10}$ & $N_{11}$
        & $P(A{=}0)$ & $P(B{=}0)$ & $\Delta x$ & $\Delta q$ \\
\midrule
$(a_0,b_0)$ & $3236$ & $252$  & $830$  & $1581$ & $0.591$ & $0.689$ & $+0.183$ & $+0.189$ \\
$(a_0,b_1)$ & $3481$ & $699$  & $602$  & $2234$ & $0.596$ & $0.582$ & $+0.192$ & $+0.082$ \\
$(a_1,b_0)$ & $3747$ & $373$  & $845$  & $1791$ & $0.610$ & $0.680$ & $+0.220$ & $+0.180$ \\
$(a_1,b_1)$ & $942$  & $3160$ & $2511$ & $549$  & $0.573$ & $0.482$ & $+0.146$ & $-0.018$ \\
\bottomrule
\end{tabular*}
\begin{minipage}{\linewidth}\vspace{4pt}\footnotesize\noindent\emph{Notes.}~
$N_{ij}$ counts postselected logical coincidences with block-$A$
outcome $i$ and block-$B$ outcome $j$, pooled over the eight CHSH
passes.  The marginals are $P(A{=}0)=(N_{00}+N_{01})/n$ and
$P(B{=}0)=(N_{00}+N_{10})/n$.  The asymmetries $\Delta x$ and
$\Delta q$ are defined in Eq.~\eqref{eq:asym}; they quantify the
departure from the balanced point at which block-$A$ readout
efficiency imbalance would cancel.
\end{minipage}
\end{table*}

\begin{table*}[tbp]
\caption{Per-pass correlators and accepted fractions by acquisition
position}
\label{tab:perpos}
\centering\footnotesize
\begin{tabular*}{\textwidth}{@{\extracolsep{\fill}}ll cccc cccc r@{}}
\toprule
 & & \multicolumn{4}{c}{Correlator $E$ by position} &
 \multicolumn{4}{c}{Accepted fraction by position} & \\
\cmidrule(lr){3-6}\cmidrule(lr){7-10}
Sess. & Pass
 & $1\,(a_0b_0)$ & $2\,(a_0b_1)$ & $3\,(a_1b_0)$ & $4\,(a_1b_1)$
 & $1\,(a_0b_0)$ & $2\,(a_0b_1)$ & $3\,(a_1b_0)$ & $4\,(a_1b_1)$
 & $n$ \\
\midrule
1 & 0 & $+0.507$ & $+0.605$ & $+0.501$ & $-0.588$ & $0.149$ & $0.167$ & $0.180$ & $0.193$ & $3447$ \\
1 & 1 & $+0.560$ & $+0.629$ & $+0.576$ & $-0.607$ & $0.149$ & $0.177$ & $0.177$ & $0.175$ & $3390$ \\
1 & 2 & $+0.656$ & $+0.638$ & $+0.684$ & $-0.563$ & $0.152$ & $0.181$ & $0.157$ & $0.176$ & $3332$ \\
1 & 3 & $+0.604$ & $+0.652$ & $+0.667$ & $-0.561$ & $0.138$ & $0.178$ & $0.176$ & $0.184$ & $3382$ \\
2 & 0 & $+0.693$ & $+0.644$ & $+0.669$ & $-0.613$ & $0.150$ & $0.172$ & $0.163$ & $0.177$ & $3307$ \\
2 & 1 & $+0.684$ & $+0.611$ & $+0.669$ & $-0.587$ & $0.149$ & $0.180$ & $0.161$ & $0.173$ & $3316$ \\
2 & 2 & $+0.654$ & $+0.656$ & $+0.684$ & $-0.558$ & $0.148$ & $0.172$ & $0.158$ & $0.176$ & $3273$ \\
2 & 3 & $+0.706$ & $+0.598$ & $+0.683$ & $-0.592$ & $0.144$ & $0.176$ & $0.179$ & $0.178$ & $3386$ \\
\midrule
\multicolumn{2}{@{}l}{Pass-avg.}
 & $+0.633$ & $+0.629$ & $+0.642$ & $-0.584$
 & $0.1475$ & $0.1754$ & $0.1689$ & $0.1790$ & --- \\
\bottomrule
\end{tabular*}
\begin{minipage}{\linewidth}\vspace{4pt}\footnotesize\noindent\emph{Notes.}~
Per-pass correlator $E$ [Eq.~\eqref{eq:E}] and accepted logical
fraction (accepted coincidences over the $5{,}000$ returned two-photon
samples of that job) at each of the four acquisition positions, with
$n$ the total accepted logical coincidences of the pass; the final row
gives the pass-averaged position means.  Within every pass the four
settings were acquired in the same fixed order, so setting and
acquisition position are perfectly confounded: each column is
therefore labeled both by its position ($1$--$4$) and by the setting
that always occupied it ($a_0b_0$, $a_0b_1$, $a_1b_0$, $a_1b_1$).
Because setting and acquisition position are one-to-one, a within-pass
temporal trend cannot be identified separately from a setting-dependent
pattern (Sec.~\ref{sec:results}).
\end{minipage}
\end{table*}

\section{Readout-imbalance sensitivity}
\label{app:readout}

The two logical output ports of block~$A$ have archived
start-of-session relative transmittances
$\kappa_{A,1}=119.99/78.75=1.524$ (session~1) and
$\kappa_{A,2}=104.98/66.15\approx1.587$ (session~2).  These ratios are
formed from the archived $t_{\rm out}$ values of
Table~\ref{tab:portmap}, which are platform-reported per-mode relative
output transmittances read from the cloud dashboard at session start,
indexed by user-facing circuit mode, with no post-compilation
re-measurement and no provider uncertainties.  Throughout this
appendix, reweighted quantities are sensitivity scenarios based on
archived session-start ratios, not calibrated corrections.  The
corresponding block-$B$ ratios are $0.994$ and $0.968$, so block~$B$
is not balanced to $1\%$ across the full acquisition; we leave it
unreweighted in the primary analysis and report a both-block
sensitivity below.  We write $\kappa_\eta$ for a generic block-$A$
efficiency ratio to reserve $\rho$ for the quantum state.  The
fair-sampling analysis of Sec.~\ref{sec:limits} leaves this imbalance
uncorrected, and we bound its effect on $S$ using the archived
session-start metadata as the primary sensitivity analysis.

Because $E$ [Eq.~\eqref{eq:E}] is a normalized ratio in which a
$b$-independent, outcome-dependent block-$A$ efficiency multiplies
numerator and denominator through the same outcome blocks, that
efficiency cancels exactly when the two block-$A$ outcomes are
balanced and equally correlated, i.e.\ $N_{00}=N_{11}$ and
$N_{01}=N_{10}$ within a setting, which is the symmetry a maximally
entangled state has in the CHSH settings.  Reweighting the outcome-$1$
counts by $\kappa_\eta$ to undo the imbalance gives the reweighted
correlator
\begin{equation}
\label{eq:Ecorr}
  E^{\rm rw} =
  \frac{(N_{00}-N_{01}) - \kappa_\eta\,(N_{10}-N_{11})}
       {(N_{00}+N_{01}) + \kappa_\eta\,(N_{10}+N_{11})} .
\end{equation}
Because $E^{\rm rw}$ is a smooth nonlinear function of the four
outcome counts of a setting, the ordinary binomial expression
$\sigma_E^2\approx(1-E^2)/n$ no longer applies once the counts are
reweighted.  We therefore obtain the shot-noise uncertainty on
$E^{\rm rw}$, and hence on $S^{\rm rw}_{\rm count}$, by the
multinomial delta method: treating the four per-setting counts as a
multinomial vector $\boldsymbol N\sim\mathrm{Mult}(n,\boldsymbol p)$
with covariance $\mathrm{Cov}(N_i,N_j)=n(p_i\delta_{ij}-p_ip_j)$, we
propagate through the gradient of Eq.~\eqref{eq:Ecorr} evaluated at
the observed cell probabilities, and combine the four independent
per-setting correlator variances into $\sigma_{S^{\rm rw}}$.  For the
session-specific primary analysis each session contributes with its
own $\kappa_{A}$, and the reported errors are conditional on those
fixed ratios.  We verified the delta-method values against a
parametric multinomial bootstrap resampling the observed four-outcome
probabilities ($10^4$ replicates per setting), which agreed to the
quoted precision; both routines are documented in the reproducibility
package.
Writing the block-$A$ outcome marginals as $N_{A0}=N_{00}+N_{01}$ and
$N_{A1}=N_{10}+N_{11}$, define the two measured asymmetries
\begin{equation}
\label{eq:asym}
  \Delta x \equiv \frac{N_{A0}-N_{A1}}{n}, \qquad
  \Delta q \equiv \frac{N_{10}-N_{11}}{n} + \frac{E}{2} ,
\end{equation}
namely the fractional imbalance of the block-$A$ outcome marginals
and the deviation of the outcome-resolved correlation
$(N_{10}-N_{11})/n$ from its balanced value $-E/2$; both vanish for
the symmetric state.  Expanding Eq.~\eqref{eq:Ecorr} about that
point,
\begin{equation}
\label{eq:Ecorr_exp}
  E^{\rm rw} \simeq E - (\kappa_\eta-1)\,(\Delta q - \tfrac12 E\,\Delta x) .
\end{equation}
The leading bias is bilinear in the efficiency mismatch
$(\kappa_\eta-1)$ and the outcome/correlation asymmetries
$(\Delta q,\Delta x)$; for $\kappa_\eta\approx1.5$--$1.6$ it is first
order in those asymmetries, not second order.

Table~\ref{tab:readout} reports the combined $S$ before and after
this reweighting.  The archived block-$A$ marginals are far from the
symmetric point: across the four settings the measured asymmetries
reach $|\Delta x| \approx 0.15$--$0.22$ and $|\Delta q|$ up to
$\approx0.19$, because the lower-efficiency outcome-$1$ port is
genuinely underpopulated.  Applying the session-specific start-of-
session ratios shifts the combined value down by $\approx0.10$, giving
the primary conditional count-pooled descriptor
\begin{equation}
\label{eq:Scorr}
  S^{\rm rw}_{\rm count} = 2.380 \pm 0.021~(\text{shot noise}),
\end{equation}
with the shot-noise error obtained by the multinomial delta method
above.  As an additional sensitivity check, applying the archived
start-of-session ratios to both blocks (scaling block-$A$
outcome-$1$ counts by $\kappa_{A}$ and block-$B$ outcome-$1$ counts
by $\kappa_{B}$) gives
\begin{equation}
\label{eq:ScorrAB}
  S^{\rm rw}_{AB} = 2.376 \pm 0.021,
\end{equation}
only $0.004$ below the block-$A$-only primary value.  We retain the
block-$A$-only result as the primary sensitivity because block~$A$
contains the dominant observed outcome imbalance; the block-$B$
ratios ($0.994$ and $0.968$) contribute a sub-percent shift under the
same fixed-ratio model.  Uniform single-kappa scans are retained as
further secondary sensitivity: $\kappa_\eta=1.5$ gives $2.386$ and
$\kappa_\eta=1.524$ gives $2.382$.

Because the platform reports no uncertainty on the transmittance
metadata, we bracket the reweighted descriptor over a range of ratios
rather than quoting the archived values as exact.  We take a
$\pm10\%$ relative excursion about the archived session ratios
($\kappa_{A,1}=1.524$, $\kappa_{A,2}=1.587$): the $+10\%$ excursion
gives $S^{\rm rw}_{\rm count} = 2.356 \pm 0.021$ and the $-10\%$
excursion gives $2.406 \pm 0.020$, so the narrow archived-metadata
band is the interval $2.356$--$2.406$ about the archived central value
$S^{\rm rw}_{\rm count} = 2.380 \pm 0.021$
[Eq.~\eqref{eq:kappa_band}].  Separately, the uniform ratio that
drives the $b_1$ remote-setting marginal difference to zero,
$\kappa_\eta = 1.395$, gives $S^{\rm rw}_{\rm count} = 2.404 \pm 0.020$,
sitting close to the $-10\%$ (upper-$S$) endpoint of this band.  This narrow band brackets
only the archived-metadata neighborhood and does not address whether a
single fixed ratio is the correct model.  The remote-setting
diagnostics of Sec.~\ref{sec:results} show that it is
mis-specified, not merely imprecise: applying the archived
session ratios flips the sign of the $b_1$ block-$B$ remote-setting
marginal difference (from $+0.100 \pm 0.008$ to $-0.033 \pm 0.009$),
overshooting zero, while leaving the $a_1$ block-$A$ difference
essentially unchanged ($+0.037 \pm 0.008 \to +0.038 \pm 0.009$).  A
constant, outcome-only reweight that both zeroes $b_1$ and leaves $a_1$
untouched therefore does not exist, so a single fixed ratio cannot be
the full model.  To bracket this model mis-specification we widen the
range to equal-session ratios $\kappa\in[1.2,1.8]$ applied to both
sessions, giving
\begin{equation}
\label{eq:kappa_band_wide_app}
  S^{\rm rw}_{\rm count} \in [\,2.341,\,2.441\,]
  \qquad(\kappa\in[1.2,1.8]),
\end{equation}
the wider mis-specification band of Eq.~\eqref{eq:kappa_band_wide}.
Both bands are statements about the assumed ratio, not about counting
statistics, and we therefore do not combine either in quadrature with
the conditional shot-noise error; the wider band is the primary systematic sensitivity
in the Abstract and Conclusion.  As an exploratory pass-level
sensitivity analysis, a random-effects meta-analysis of the eight
session-specifically reweighted pass values (Table~\ref{tab:chsh})
gives $\tau^2 = 0.024$ ($\tau = 0.15$), pooled estimate
\begin{equation}
\label{eq:result_corr_pass}
  S^{\rm rw}_{\rm pass} = 2.383, \qquad
  [\,2.242,\ 2.524\,]\ \ (t_7),
\end{equation}
Hartung--Knapp standard error $0.060$ under an exchangeability working
model.  The residual
marginal diagnostics of Sec.~\ref{sec:results} indicate that the
actual outcome-dependent filtering may be more complicated than
session-fixed efficiency ratios.  Under an equal-session uniform-ratio
scan (Table~\ref{tab:skappa}),
$S^{\rm rw}_{\rm count}$ decreases monotonically from $2.485$ at
$\kappa_\eta = 1$ to $2.315$ at $\kappa_\eta = 2$; the archived
session-specific pair ($\kappa_{A,1}=1.524$, $\kappa_{A,2}=1.587$)
gives $S=2.3796 \pm 0.0207$.  Extreme ratios ($\kappa_\eta \approx 12$
and the $\kappa_\eta \to \infty$ limit $1.89$) are tabulated for
completeness.

\begin{table}[tbp]
\caption{Equal-session $S(\kappa)$ scan of the count-pooled reweighted
score}
\label{tab:skappa}
\centering\footnotesize
\setlength{\tabcolsep}{6pt}
\begin{tabular}{@{}cc@{\hskip 18pt}cc@{}}
\toprule
$\kappa$ & $S^{\rm rw}_{\rm count}$ & $\kappa$ & $S^{\rm rw}_{\rm count}$ \\
\midrule
$1.000$ & $2.4854(192)$ & $1.700$ & $2.3550(211)$ \\
$1.100$ & $2.4625(194)$ & $1.800$ & $2.3409(214)$ \\
$1.200$ & $2.4413(197)$ & $2.000$ & $2.3151(219)$ \\
$1.300$ & $2.4216(200)$ & $3.000$ & $2.2202(240)$ \\
$1.395$ & $2.4041(203)$ & $4.000$ & $2.1598(255)$ \\
$1.500$ & $2.3861(206)$ & $6.000$ & $2.0872(275)$ \\
$1.587$ & $2.3721(208)$ & $12.000$ & $1.9983(301)$ \\
\bottomrule
\end{tabular}
\begin{minipage}{\linewidth}\vspace{4pt}\footnotesize\noindent\emph{Notes.}~
Count-pooled reweighted score $S^{\rm rw}_{\rm count}$ as a function of
a single equal-session ratio $\kappa$ applied to the block-$A$
outcome-$1$ counts of both sessions [Eq.~\eqref{eq:Ecorr}]; parenthetic
figures are the last-digit shot-noise uncertainties from the
multinomial delta method (e.g.\ $2.4854(192)=2.4854\pm0.0192$).  This
is an \emph{ad hoc} stress scan, not a calibrated uncertainty band;
the archived session-specific value ($\kappa_{A,1}=1.524$,
$\kappa_{A,2}=1.587$) is $S=2.3796\pm0.0207$.
\end{minipage}
\end{table}

The per-pass and per-session
reweighted values under the primary session-specific calibration are
listed in Table~\ref{tab:chsh}; the weakest reweighted pass
($2.040 \pm 0.062$) overlaps $2$ within $1\sigma$.  Each ratio was recorded at the corresponding session
start and was not re-measured during that session, so its temporal
stability across passes, settings, and recompilations is assumed
rather than demonstrated.
The reweighting script is included in the
reproducibility package, and the per-setting asymmetries used by the
reweight are tabulated in Table~\ref{tab:marginals}.

\begin{table}[tbp]
\caption{Sensitivity to readout-port imbalance}
\label{tab:readout}
\begin{tabular}{@{}lc@{}}
\toprule
Estimate & $S$ \\
\midrule
Uncorrected (fair sampling) & $2.485 \pm 0.019$ \\
Session-specific block-$A$ reweight & $2.380 \pm 0.021$ \\
\quad narrow $\kappa$ metadata band ($\pm10\%$) & $2.356$--$2.406$ \\
\quad wide $\kappa$ mis-spec.\ band ($\kappa\in[1.2,1.8]$) & $2.341$--$2.441$ \\
Both-block session-specific reweight & $2.376 \pm 0.021$ \\
Marginal-consistent ($\kappa_\eta = 1.395$) & $2.404 \pm 0.020$ \\
Uniform reweight ($\kappa_\eta = 1.5$) & $2.386 \pm 0.021$ \\
Uniform reweight ($\kappa_\eta = 1.524$) & $2.382 \pm 0.021$ \\
\bottomrule
\end{tabular}
\begin{minipage}{\linewidth}\vspace{4pt}\footnotesize\noindent\emph{Notes.}~
Combined CHSH value before and after reweighting accepted counts by
archived start-of-session output-transmittance ratios.  The primary
block-$A$ row uses $\kappa_{A,1}=1.524$ and $\kappa_{A,2}=1.587$
[Eq.~\eqref{eq:Ecorr}]; the both-block row also applies
$\kappa_{B,1}=0.994$ and $\kappa_{B,2}=0.968$ to block-$B$
outcome-$1$ counts.  The narrow $\kappa$ metadata band is a $\pm10\%$
relative excursion about the archived session ratios
[Eq.~\eqref{eq:kappa_band}], with the marginal-consistent ratio
$1.395$ ($S=2.404$) sitting near its upper-$S$ endpoint; the wide
mis-specification band spans equal-session ratios $\kappa\in[1.2,1.8]$
[Eq.~\eqref{eq:kappa_band_wide}]; it is an \emph{ad hoc} stress scan
rather than a calibrated uncertainty band, and it is the leading
systematic on the reweighted score.  Neither band is combined in
quadrature with the shot noise.  Uniform
single-$\kappa_\eta$ rows are secondary
scans of the block-$A$ channel alone.  Uncertainties are conditional
count-pooled shot noise obtained by the multinomial delta method; the
separate session-specific pass-level estimator is
$S^{\rm rw}_{\rm pass} = 2.383$ with Hartung--Knapp standard error
$0.060$ [Eq.~\eqref{eq:result_corr_pass}].
\end{minipage}
\end{table}

\end{document}